\documentclass[fleqn,usenatbib]{mnras}

\usepackage{newtxtext,newtxmath}

\usepackage[T1]{fontenc}
\usepackage{ae,aecompl}

\usepackage{graphicx}	
\usepackage{amsmath}	
\usepackage{slashed}    
\DeclareMathOperator*{\argminT}{argmin}
\DeclareMathOperator*{\argmaxT}{argmax}
\DeclareMathOperator*{\minT}{min}
\DeclareMathOperator*{\maxT}{max}
\usepackage{mathtools}
\DeclarePairedDelimiterX{\norm}[1]{\lVert}{\rVert}{#1}
\makeatletter
\newcommand*{\rom}[1]{\expandafter\@slowromancap\romannumeral #1@}
\makeatother
\usepackage{wrapfig}
\usepackage{bm}
\usepackage{dsfont}
\usepackage[super]{nth}
\usepackage[utf8]{inputenc}

\usepackage{smartdiagram}
\usepackage{tikz}
\usetikzlibrary{shapes,arrows}
\tikzstyle{block} = [rectangle, draw, fill=yellow!10,
    text width=12em, text centered, rounded corners, minimum height=3em]
\tikzstyle{line} = [draw, -latex']
\tikzstyle{decision} = [rectangle, draw, fill=green!10, text width=12em, text centered, rounded corners, minimum height=3em]
\tikzstyle{parameter_fixed} = [rectangle, draw, fill=purple!10, text width=12em, text centered, rounded corners, minimum height=3em]
\tikzstyle{iteration} = [rectangle, draw, fill=blue!10, text width=12em, text centered, rounded corners, minimum height=3em]

\newcommand{\emdash}{\nobreak--\nobreak\hskip4pt}

\usepackage{algorithm}
\usepackage{setspace}
\usepackage[noend]{algpseudocode}
\makeatletter
\def\BState{\State\hskip-\ALG@thistlm}
\makeatother

\title[Mass-mapping: peak statistics and feature locations]{Sparse Bayesian mass-mapping with uncertainties: peak statistics and feature locations}

\author[Price et al.]{
M.~A.~Price$^{1}$\thanks{E-mail: m.price.17@ucl.ac.uk}, J.~D.~McEwen$^{1}$, X.~Cai$^{1}$, T.~D.~Kitching$^{1}$ \newauthor
\normalsize(for the LSST Dark Energy Science Collaboration)\\
$^{1}$Mullard Space Science Laboratory, University College London, RH5 6NT, UK.
}

\date{Accepted XXX. Received YYY; in original form ZZZ}

\pubyear{2018}

\begin{document}
\label{firstpage}
\pagerange{\pageref{firstpage}--\pageref{lastpage}}
\maketitle

\begin{abstract}
Weak lensing convergence maps \emdash upon which higher order statistics can be calculated \emdash can be recovered from observations of the shear field by solving the lensing inverse problem. For typical surveys this inverse problem is ill-posed (often seriously) leading to substantial uncertainty on the recovered convergence maps. In this paper we propose novel methods for quantifying the Bayesian uncertainty in the location of recovered features and the uncertainty in the cumulative peak statistic \emdash the peak count as a function of signal to noise ratio (SNR). We adopt the sparse hierarchical Bayesian mass-mapping framework developed in previous work, which provides robust reconstructions and principled statistical interpretation of reconstructed convergence maps without the need to assume or impose Gaussianity. We demonstrate our uncertainty quantification techniques on both Bolshoi N-body (cluster scale) and Buzzard V-1.6 (large scale structure) N-body simulations. For the first time, this methodology allows one to recover approximate Bayesian upper and lower limits on the cumulative peak statistic at well defined confidence levels.
\end{abstract}



\section{Introduction}

In an empty universe the \textit{null geodesics} along which photons travel correspond directly to straight lines. However, in the presence of a non-uniform distribution of matter the null geodesics are perturbed \textit{via} gravitational interaction with the local matter over or under density \textit{i.e.} the photons are \textit{gravitationally lensed} \citep{[48], [2], [23], [24]}. As this gravitational interaction is sensitive only to the total matter distribution, and the overwhelming majority of matter is typically dark, gravitational lensing provides a natural probe of dark matter itself \citep{[25]}.
\par
Collections of associated photons emitted from a distant object travel along separate geodesics which are perturbed in different ways by the intervening matter distribution, \textit{e.g.} photons traveling closer to matter over densities will interact more strongly and therefore be perturbed more than those farther away. As such the geometry of a distant object is warped \citep{[1]} -- \textit{i.e.} colloquially the object is \textit{lensed}.
\par
Provided the propagating photons at no time come closer than one Einstein radius to the intervening matter over and under densities, the object is \textit{weakly lensed}. Weak gravitational lensing of distant galaxies manifests itself at first order into two quantities; the spin-0 convergence $\kappa$ which is a magnification, and the spin-2 shear $\gamma$ which is a perturbation to the galaxy ellipticity (third-flattening).
\par
Both the shear $\gamma$ and the convergence $\kappa$ have dominant intrinsic (\textit{i.e.} in the absence of lensing effects) underlying values which makes measurements of the lensing effect difficult. In fact, there is no \textit{a priori} way to know the intrinsic brightness of a galaxy (resulting in an inherent degeneracy -- the mass-sheet degeneracy) and so the convergence is not an observable quantity. In fact, the standard convergence is not gauge invariant and is therefore fundamentally unobservable \citep{[48]}. However, as the intrinsic ellipticity distribution of galaxies has zero mean one can average to recover the shearing contribution, hence the shear is an observable quantity. As such, measurements of the shear field are taken and inverted to form estimators of the convergence. Typically this inverse problem is seriously ill-posed and so substantial uncertainty on the reconstructed convergence map is introduced.
\par
A wealth of information may be calculated directly from the shear field \citep[often in the form of second order statistics \citep{kilbinger2015} -- such as the power spectrum as in][]{[26],[56]} though recently there is increasing interested in extracting Non-Gaussian information from the convergence field, \textit{e.g.} peak statistics, Minkowski functionals, extreme value statistics \citep{[27], [28], [52], [54], [55]}.
\par
Primarily, the interest has arisen as higher-order statistics of the convergence field have been shown to provide complementary constraints on dark matter cosmological parameters which are typically poorly constrained by second-order statistics \citep{[50]}.
\par
However, to make principled statistical inferences from the convergence field, the inversion from $\gamma$ to $\kappa$ must be treated in a principled statistical manner \emdash something which until recently was missing from convergence reconstruction algorithms which were either not framed in a statistical framework \citep[\textit{e.g.}][]{[29],[5],[6],[3],[15]} or made assumptions of Gaussianity \citep[\textit{e.g.}][]{[30],[26],[43]}. As the information of interest in higher-order convergence statistics is non-Gaussian, assumptions of Gaussianity in the reconstruction process severely degrade the quality of the cosmological information.
\par
Recently, a mass-mapping framework was developed \citep[see][]{[M1]} which addressed precisely this issue. This new sparse hierarchical Bayesian mass-mapping formalism can be rapidly computed, can be extended to big data, and provides a principled statistical framework for quantifying uncertainties on reconstructed convergence maps. Notably, it has been shown to accurately reconstruct very high dimensional Bayesian estimators many orders of magnitude faster than \textit{state-of-the-art} proximal MCMC algorithms \emdash it was specifically benchmarked against Px-MALA \citep{[51],Durmus2018} in \citet{[M2]}.
\par
In this paper, we propose two novel uncertainty quantification techniques, aimed to answer two frequently asked questions of the recovered convergence map. The first of these questions asks where a feature of interest in the reconstructed convergence map could have been observed \emdash typically this has been addressed by bootstrapping; however we can now infer it directly in a Bayesian manner. The second question asks given a magnitude threshold what is the maximum and minimum number of peaks which could have been observed, within some well defined confidence.
\par
The structure of this article is as follows. To begin, in section \ref{sec:WeakGravitationLensing} we provide cursory introduction to weak lensing from a mathematical perspective, with emphasis on the weak lensing planar forward model in subsection \ref{sec:LensingForwardModel}. Following this we provide a brief overview of Bayesian inference and the previously developed \citep{[M1]} sparse hierarchical Bayesian mass-mapping algorithm in section \ref{sec:VERITAS}. An introduction to Bayesian credible regions, specifically the highest posterior density credible region is provided in section \ref{sec:HPD_region}. In section \ref{sec:BayesianLocation} we develop a novel Bayesian inference approach to quantifying the uncertainty in reconstructed feature location, which we then showcase on illustrative N-body cluster simulation data in section \ref{sec:BayesLocationApplication}. We then introduce a novel Bayesian inference approach for recovery of principled uncertainties on the aggregate peak count statistic in section \ref{sec:peak_uncertainties}. Following this we showcase this Bayesian inference approach to quantify uncertainty in the aggregate peak statistic in section \ref{sec:PeakDemonstration} on N-body large scale structure (LSS) illustrative simulation data. Finally we draw conclusions in section \ref{sec:Conclusion}.

\section{Weak Gravitational Lensing} \label{sec:WeakGravitationLensing}
In this section we provide a brief introduction to weak gravitational lensing, with emphasis on how this effect manifests itself into observable quantities. For a detailed background review of the field see \citet{[1],[2]}. For a more mathematical background of the field, with emphasis on statistical methods see \citet{[48], [23], [24]}. For a background of specifically the peak statistics see \citet{[53]}.
\subsection{Mathematical Background} \label{sec:MathematicalBackground}
In a non-uniform distribution of matter the null geodesics along which photons travel are no longer straight lines, instead they are now sensitive to the local matter distribution. When many photons are propagating from a distant object to us here and now, the local matter distribution adjusts the geometry of the object we observe \emdash the object is \textit{gravitationally lensed}.
\par
Provided the trajectory of the propagating photons at no time comes closer than one Einstein radius $\theta_E$ to the intervening matter over densities then the lens equation,
\begin{equation}
\beta = \theta - \theta_E^2 \frac{\theta}{|\theta|^2},
\end{equation}
remains effectively singular and we are in the \textit{weak lensing} regime. Equivalently one can define the weak lensing regime to be convergence fields for which $\kappa \ll 1$ -- ensuring that the shear signal remains linear. Here the Einstein radius is given by,
\begin{equation}
\theta_E = \sqrt{\frac{4GM_{\text{lens}}}{c^2}\frac{f_K(r - r^{\prime})}{f_K(r)f_K(r^{\prime})}},
\end{equation}
where $G$ is the gravitational constant, $M_{\text{lens}}$ is lensing mass, $c$ is the speed of light \textit{in vacuo} and $f_K(r)$ is the angular diameter distance defined as:
\begin{equation} \label{eq:angular_diameter_distance}
  f_K(r)=\begin{cases}
               \sin(r) \quad &\text{if}\quad  K = 1, \\
               r \quad &\text{if}\quad K = 0,\\
               \sinh(r) \quad &\text{if}\quad K = -1,\\
            \end{cases}
\end{equation}
where $r$ is the comoving distance and $K$ is the curvature of the universe, which has been observed to be consistent with $0$ by \citet{[47]}.
\par
As galaxies are naturally sparsely distributed across the sky, most observations fall within the weak lensing regime. The weak gravitational lensing effect can be described by a lensing potential $\phi(r,\omega)$ which is the integrated Newtonian potential $\Phi(r, \omega)$ along the line of sight
\begin{equation} \label{eq:lineofsightmass}
\phi(r,\omega) = \frac{2}{c^2} \int_0^r dr^{\prime} \frac{f_K(r - r^{\prime})}{f_K(r) f_K(r^{\prime})} \Phi(r^{\prime}, \omega),
\end{equation}
where $\omega = (\theta, \psi)$ are angular spherical co-ordinates. A further constraint exists, such that the local Newtonian potential $\Phi(r,\omega)$ must satisfy the Poisson equation:
\begin{equation} \label{eq:Poisson}
\nabla^2 \Phi(r,\omega) = \frac{3 \Omega_M H_0^2}{2a(r)} \delta(r,\omega),
\end{equation}
for matter density parameter $\Omega_M$, Hubble constant $H_0$ and scale parameter $a(r)$. Combined, equations (\ref{eq:lineofsightmass}) and (\ref{eq:Poisson}) allow one to make inferences of cosmological parameters from observations of the lensing potential $\phi(r,\omega)$.
\par
At linear order, gravitational lensing manifests itself as two quantities: the spin-0 convergence field $\kappa$ (magnification) and the spin-2 shear field $\gamma$ (perturbation to ellipticity). It can be shown that \citep{[1],[2]} these quantities can be related to the lensing potential $\phi(r,\omega)$ by,
\begin{align}
& \kappa(r,\omega) = \frac{1}{4}(\eth \bar{\eth} + \bar{\eth} \eth) \; \phi(r,\omega), \label{eq:kappatophi} \\
& \gamma(r,\omega) = \frac{1}{2} \eth \eth \; \phi(r,\omega), \label{eq:gammatophi}
\end{align}
where $\eth$ and $\bar{\eth}$ are the spin raising and lowering operators respectively and are in general defined to be:
\begin{align}
\eth \equiv -\sin^s\theta \Big ( \frac{\partial}{\partial \theta} + \frac{i \partial}{\sin\theta \partial \psi} \Big ) \sin^{-s}\theta, \\
\bar{\eth} \equiv -\sin^{-s}\theta \Big ( \frac{\partial}{\partial \theta} - \frac{i \partial}{\sin\theta \partial \psi} \Big ) \sin^{s}\theta.
\end{align}

\subsection{Lensing Planar Forward Model} \label{sec:LensingForwardModel}
Often second order statistics \citep{kilbinger2015} related to the shear $\gamma$ are computed \citep[\textit{e.g.} the shear power spectrum as in][]{[56],[26]}, though increasingly there is interest in extracting weak lensing information from the convergence directly \emdash typically higher-order non-Gaussian information.
\par
Unfortunately, due to an inherent degeneracy the convergence is not an observable quantity \citep{[1],[2],[48]} \emdash this effect is colloquially referred to as the \textit{mass-sheet degeneracy}. However, as the intrinsic ellipticity distribution of galaxies has zero mean, averaging many ellipticity observations within a given pixel provides a good estimate of the shear field.
\par
In fact, there exists a further degeneracy between $\kappa$ and $\gamma$ such that the true observable is the \textit{reduced shear} $g$ but for the context of the paper we will assume $\gamma \approx g \ll 1$ \emdash see \citeauthor{[57]} pg.153, \citet{[M1]}, or \citet{[3]} for details on how to account for the non-linear reduced shear.
\par
As both $\kappa$ and $\gamma$ are related to $\phi$ a relation between $\kappa$ and $\gamma$ can be formed. Therefore, typically measurements of the shear field are taken and inverted to form estimates of the underlying convergence field. For small fields of view the \textit{flat sky approximation} can be made, which reduces the spin-raising $\eth$ and lowering $\bar{\eth}$ operators to \citep{[4]}:
\begin{equation}
\eth \approx -(\partial_x + i \partial_y) \quad \mbox{and} \quad \bar{\eth} \approx -(\partial_x - i \partial_y).
\end{equation}
From equations (\ref{eq:kappatophi}) and (\ref{eq:gammatophi}) it is clear that the forward model in Fourier space between $\kappa$ and $\gamma$ is given by
\begin{equation} \label{eq:forwardmodel}
\hat{\gamma}(k_x,k_y) = \bm{\mathsf{D}}_{k_x,k_y} \hat{\kappa}(k_x, k_y),
\end{equation}
with the mapping operator being
\begin{equation} \label{eq:forward_model}
\bm{\mathsf{D}}_{k_x,k_y} = \frac{k_x^2-k_y^2+2ik_xk_y}{k_x^2+k_y^2},
\end{equation}
where we have dropped the spin subscripts for clarity. To recover an estimator of the convergence one must invert this forward model.
\par
The most naive inversion technique is that of Kaiser-Squires (KS) inversion \citep{[5]}, which is direct inversion in Fourier space, \textit{i.e.}
\begin{equation} \label{eq:Kaiser-Squires}
\hat{\kappa}^{\text{KS}} = \bm{\mathsf{D}}^{-1} \hat{\gamma},
\end{equation}
where we have again dropped function arguments and subscripts for clarity. This approach attempts to mitigate the effect of noise by convolving the recovered convergence estimate with a broad Gaussian smoothing kernel, which severely degrades the quality of non-Gaussian information. This poses a somewhat serious issue as non-Gaussian information is precisely the information that is to be extracted from the convergence field. Therefore for higher-order convergence statistics the KS estimator is patently sub-optimal.
\par
Moreover, decomposition of spin fields on bounded manifolds is well known to be degenerate \citep{[4]} and so inversion of shear to convergence for masked fields is inherently ill-defined \emdash in particular the KS estimator is known to break down for non-trivial masking. In fact the lensing inverse problem is often seriously ill-posed, therefore motivating methods regularized by prior information.

\section{Sparse Bayesian Mass-mapping} \label{sec:VERITAS}
Many mass-mapping algorithms have been considered \citep[\textit{e.g.}][]{[29],[5],[6],[3],[15],[49],[22]}, however in the context of this paper we wish to conduct principled statistical analysis of the reconstructed convergence map, and so we opt for the sparse hierarchical Bayesian algorithm presented in \citet{[M1]} and benchmarked against MCMC algorithms in \citet{[M2]}.
\par
Recently a sparse hierarchical Bayesian framework for convergence reconstruction was presented \citep{[M1]} which is not limited to Gaussian priors \emdash in fact the prior need not even be differentiable. In this work we adopt this mass-mapping algorithm, which we briefly describe below.
\par
First, by Bayes' theorem the \textit{posterior distribution} of the convergence $\kappa$ reads
\begin{equation} \label{eq:bayes}
p(\kappa|\gamma) = \frac{p(\gamma|\kappa)p(\kappa)}{\int_{\mathbb{C}^N} p(\gamma|\kappa)p(\kappa)d\kappa},
\end{equation}
which shows how one should infer the \textit{posterior} $p(\kappa|\gamma)$ from the \textit{likelihood function} (data fidelity term)  $p(\gamma|\kappa)$ and the \textit{prior} (regularization term) $p(\kappa)$ \citep[see \textit{e.g.}][for a clear introduction to Bayesian inference in a cosmological setting]{[46]}. In the scope of this paper we do not consider the \textit{Bayesian evidence} $\int_{\mathbb{C}^N} p(\gamma|\kappa)p(\kappa)d\kappa$ as it acts as a normalization term and so does not effect the recovered solution. Typically the Bayesian evidence is used for model comparison which is an avenue of study in of itself.
\par
In a Bayesian inference problem one often wishes to find the solution $\kappa$ which maximizes the posterior given data and some model. From the monotonicity of the logarithm function, maximization of the posterior is equivalent to minimization of the \textit{log-posterior} such that
\begin{equation} \label{eq:log-posterior}
\argmaxT_{\kappa} \big \lbrace p(\kappa|\gamma) \big \rbrace \equiv \argminT_{\kappa} \big \lbrace -\log ( \; p(\kappa|\gamma) \;) \big \rbrace,
\end{equation}
where the convergence $\kappa$ which maximizes the posterior is given by $\kappa^{\text{map}}$, where MAP stands for \textit{maximum a posteriori}. Provided the posterior is log-concave the minimization of the log-posterior takes the form of a \textit{convex optimization} problem, which can be rapidly computed even in high dimensional settings.
\par
Let $\kappa \in \mathbb{C}^N$ be the discretized complex convergence field and $\gamma \in \mathbb{C}^M$ be the discretized complex shear field, where $M$ is the number of binned shear measurements and $N$ is the dimensionality of the recovered convergence field. Suppose we can define a \textit{measurement operator} $\bm{\Phi} \in \mathbb{C}^{M \times N}: \kappa \mapsto \gamma$ which maps $\kappa$ onto $\gamma$. On the plane, the measurement operator is given by \citep[\textit{e.g.}][]{[M1]}
\begin{equation} \label{eq:measurement_operator}
\bm{\Phi} = \bm{\mathsf{F}}^{-1} \bm{\mathsf{D}} \bm{\mathsf{F}},
\end{equation}
where $\bm{\mathsf{D}}$ is the planar forward model in Fourier space as defined in equation (\ref{eq:forward_model}), and $\bm{\mathsf{F}}$ $(\bm{\mathsf{F}}^{-1})$ is the forward (inverse) fast Fourier transform.
\par
Now suppose that some noise $n$ is contaminating our measurements, then measurements are obtained through the \textit{measurement equation}
\begin{equation}
\gamma = \bm{\Phi} \kappa + n.
\end{equation}
Within this article we will consider the typical case in which the noise $n \sim \mathcal{N}(0,\sigma_n^2) \in \mathbb{C}^M$, \textit{i.e.} i.i.d. (independent and identically distributed) zero mean Gaussian noise of variance $\sigma^2_n$. In this setting the likelihood function is given by a multivariate Gaussian, which for diagonal covariance $\Sigma = \sigma_n^2 \mathbb{I}$ is given compactly as
\begin{equation}
p(\gamma|\kappa) \propto \exp \Bigg(\frac{-\norm{\bm{\Phi} \kappa - \gamma}_2^2}{2\sigma_n^2} \Bigg),
\end{equation}
which \citep[as in][]{[M1]} is regularized by a sparsity promoting Laplace-type $l_1$-norm wavelet prior
\begin{equation}
p(\kappa) \propto \exp \Big(-\mu \norm{\bm{\Psi}^{\dag}\kappa}_1 \Big),
\end{equation}
where $\mu \in \mathbb{R}_{+}$ is the jointly inferred MAP regularization parameter \citep{[16]} \emdash the derivation and implementation of which may be found in \citet{[M1]}. Sparsity promoting priors in wavelet dictionaries have explicitly been shown to be effective in the weak lensing setting \citep{[15], [6], [9], [M1]}.
\par
Using these terms, the minimization of the log-posterior becomes
\begin{equation} \label{eq:optimization}
\kappa^{\text{map}} = \argminT_{\kappa} \Bigg \lbrace \mu \norm{ \bm{\Psi}^{\dag}\kappa}_1 + \frac{\norm{\bm{\Phi} \kappa - \gamma}_2^2}{2\sigma_n^2} \Bigg \rbrace,
\end{equation}
which is then solved iteratively by implementing a proximal forward-backward splitting algorithm \citep[\textit{e.g.}][]{[40]}.
\par
Note that one can choose any convex log-priors \textit{e.g.} an $\ell_2$-norm prior from which one essentially recovers Weiner filtering \citep[see][for alternate iterative Weiner filtering approaches]{Seljak2003,Horowitz2018}.

\begin{figure*}
\begin{center}
\begin{tikzpicture}[node distance = 2cm, auto]
    \node [parameter_fixed, text width=14em] (-1) {Calculate MAP solution: $\kappa^{\text{map}}$ (\ref{eq:optimization})};
    \node [below of=-1,node distance = 3cm](0) {\includegraphics[width=0.25\textwidth]{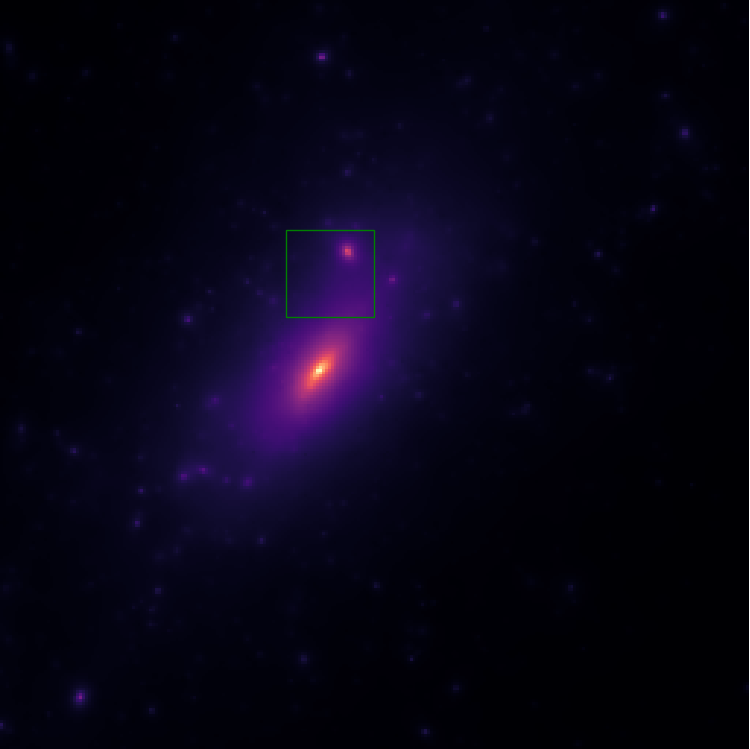}};
    \node [below of=-1, node distance=1.6cm, text width=18em] (dummy) {};
    \node [block, left of=dummy, node distance=5.2cm, text width=14em] (1a) {Remove feature $\mathcal{Z}$ by equation (\ref{eq:inpainting}).};
    \node [block, right of=dummy, node distance=5.2cm, text width=14em] (1b) {Extract feature $\mathcal{Z} = \kappa^{\text{map}} \mathbb{I}_{\Omega_{\mathcal{Z}}}$};
    \node [below of=1a,node distance = 3cm](2a) {\includegraphics[width=0.25\textwidth]{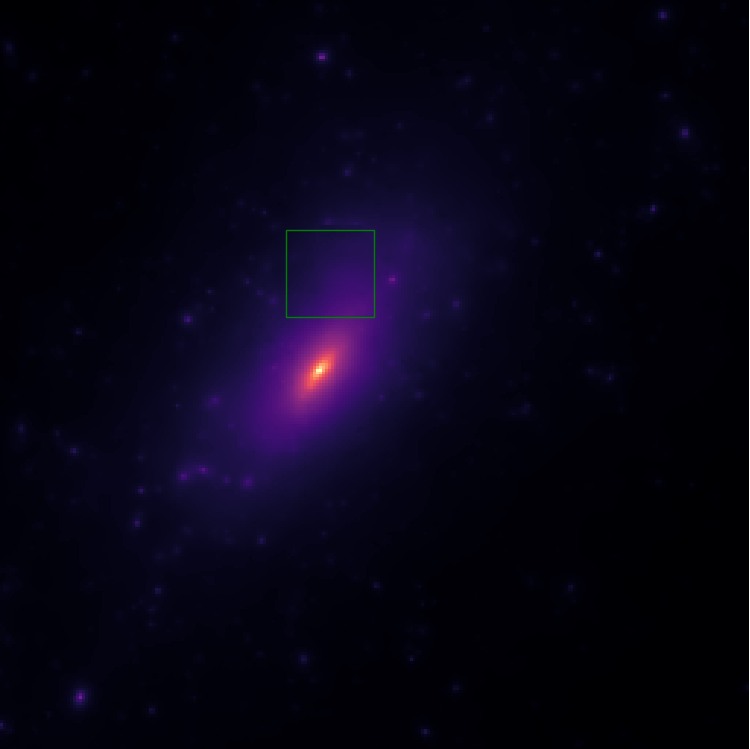}};
    \node [below of=1b,node distance = 3cm](2b) {\includegraphics[width=0.25\textwidth]{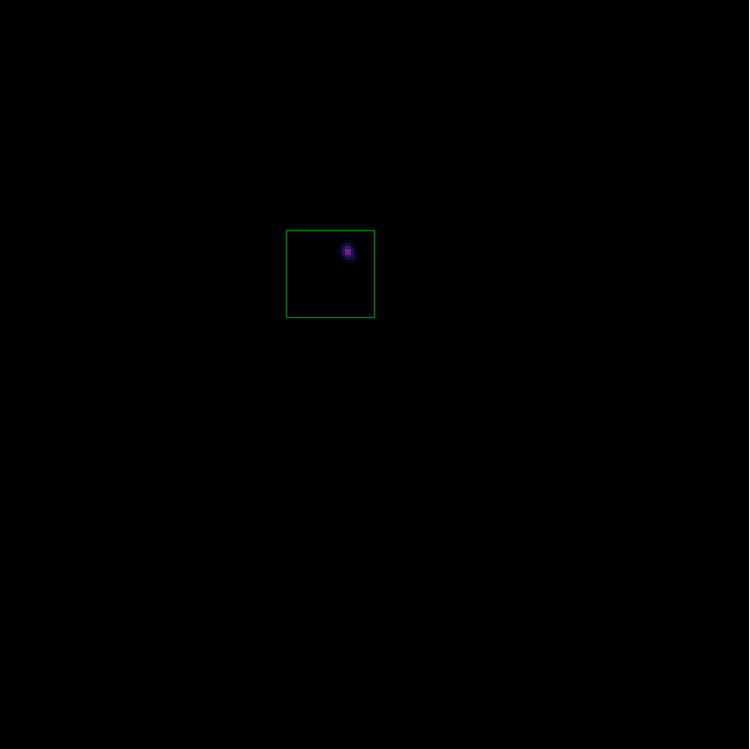}};
    \node [block, below of=0, node distance=4.0cm, text width=14em] (3) {Insert: feature $\mathcal{Z}$ at $\bm{x}_t$, get surrogate $\kappa^{\text{sgt}}$};
    \node [below of=3,node distance = 3cm](4) {\includegraphics[width=0.25\textwidth]{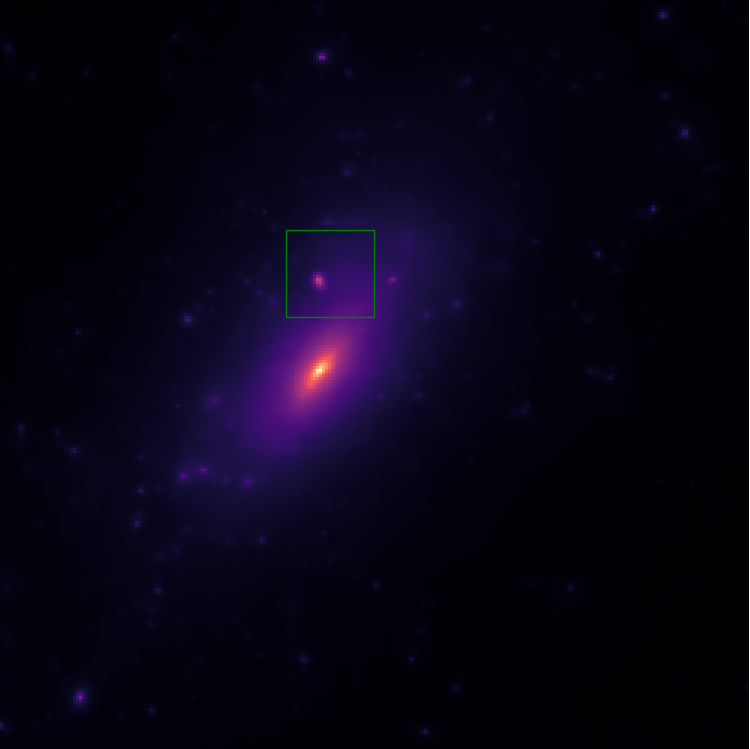}};
    \node [decision, below of=4, node distance=3.0cm, text width=14em] (5) {Is: $\kappa^{\text{sgt}} \in C_{\alpha}^{\prime}$ ? };
    \node [iteration, below of=2b, node distance=7.0cm, text width=8em] (7) {Reject pixel: $\bm{x}_t$};
    \node [iteration, below of=2a, node distance=7.0cm, text width=8em] (8) {Accept pixel: $\bm{x}_t$};
    \node [iteration, below of=2a, node distance=4.0cm, text width=8em] (9) {$t \rightarrow t + 1$};
    \node [iteration, below of=2b, node distance=4.0cm, text width=8em] (10) {$t \rightarrow t + 1$};

    \path [line] (-1) -- (0);
    \path [line] (0) -- (1a);
    \path [line] (0) -- (1b);
    \path [line] (1a) -- (2a);
    \path [line] (1b) -- (2b);
    \path [line] (2a) -- (3);
    \path [line] (2b) -- (3);
    \path [line] (3) -- (4);
    \path [line] (4) -- (5);
    \path [line,dashed] (5) -- node[below]{Yes}(8);
    \path [line,dashed] (8) -- node[decision,left]{Select next nearest pixel.}(9);
    \path [line,dashed] (9) -- (3);
    \path [line,dashed] (10) -- (3);
    \path [line,dashed] (7) -- node[decision,right]{Select next nearest pixel.}(10);
    \path [line,dashed] (5) -- node[below]{No}(7);

\end{tikzpicture}
\caption{Schematic representation of the inverse nested iterations to determine the \textit{Bayesian location} (see section \ref{sec:BayesianLocation}). The Bayesian location is a positional uncertainty on a feature of interest $\mathcal{Z}$ within a recovered convergence field. Once a complete ring of pixels have been rejected the algorithm returns a binary map of accepted pixels which we call the Bayesian location. Any pixel outside of this location is rejected at $100(1-\alpha)\%$ confidence. Alternately the probability isocontour bounding the set of acceptable pixels can be located by N-splitting circular bisection as described in section \ref{sec:N-splitting} and Appendix \ref{sec:appendixa}.}
\label{fig:BayesianLocationSchema}
\end{center}
\end{figure*}
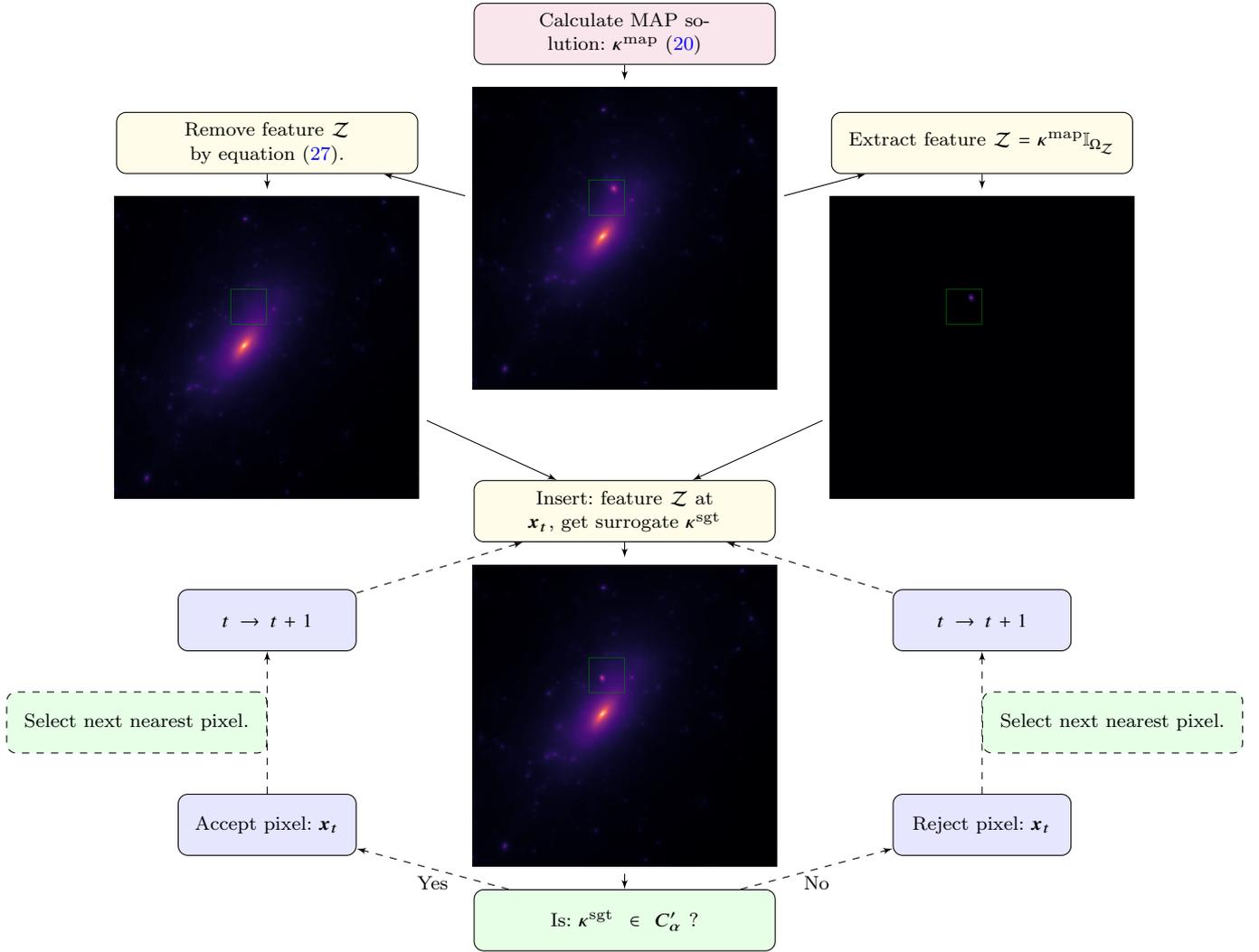

\begin{figure}
	\centering
    \large Bayesian Location of Bolshoi-3 Sub-halos \par
	\includegraphics[width=0.45\textwidth, trim={0 0 0 1.6cm},clip]{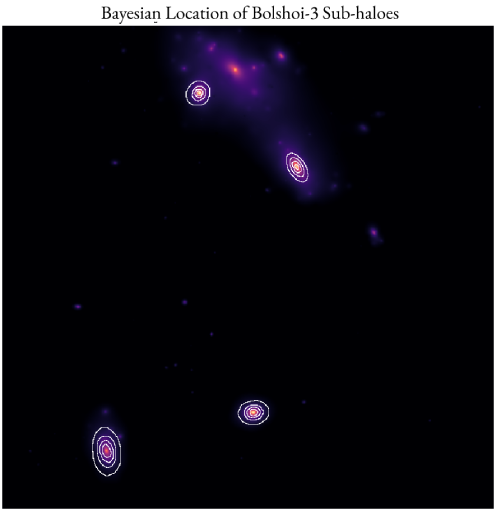}
    \caption{Combined plot of the $99\%$ confidence Bayesian locations at SNR $= 12, 15, 17, 20$ dB. The outer rings represent the noiser position isocontours whereas as the data becomes cleaner the isocontour ring becomes smaller (therefore the rings represent isocontours at SNR $= 12, 15, 17, 20$ dB, from the outer rings inwards respectively). N-splitting Circular Bisection (see section \ref{sec:N-splitting}) was used to efficiently compute each isocontour. For input SNR's below $\approx 10$ the smaller local features cannot be determined physical \textit{via} the initial hypothesis test, and so we truncate our analysis at SNR = 12.}
    \label{fig:Combined_peaks}
\end{figure}

\subsection{Bayesian credible regions} \label{sec:HPD_region}
In Bayesian analysis a posterior credible region $C_{\alpha} \in \mathbb{C}^N$ at confidence $100(1-\alpha)\%$ is a set which satisfies:
\begin{equation} \label{eq:CredibleIntegral}
p(\kappa \in C_{\alpha}|\gamma) = \int_{\kappa \in \mathbb{C}^N} p(\kappa|\gamma)\mathbb{I}_{C_{\alpha}}d\kappa = 1 - \alpha,
\end{equation}
where $\mathbb{I}_{C_{\alpha}}$ is an indicator function defined such that,
\begin{equation} \label{eq:indicator}
  \mathbb{I}_{C_{\alpha}}=\begin{cases}
               1 \quad \text{if}\quad  \kappa \in C_{\alpha}, \\
               0 \quad \text{if}\quad \kappa \not\in C_{\alpha}.\\
            \end{cases}
\end{equation}
\par
There are in general a large number of posterior regions (hyper-volumes) which satisfy this integral. The decision-theoretically optimal region \emdash in the sense of minimum enclosed volume \emdash is called the \textit{highest posterior density} (HPD) credible region \citep{[19]} and is defined to be:
\begin{equation}
C_{\alpha} := \lbrace \kappa : f(\kappa) + g(\kappa) \leq \epsilon_{\alpha} \rbrace,
\end{equation}
where $f(\kappa) = \mu \norm{\Psi^{\dag} \kappa}_1$ is the log-prior term and $g(\kappa) = \norm{\Phi \kappa - \gamma}_2^2 / 2\sigma_n^2$ is our data fidelity term (log-likelihood function). Here $\epsilon_{\alpha}$ defines a level-set (\textit{i.e.} isocontour) of the log-posterior set such that equation (\ref{eq:CredibleIntegral}) is satisfied. In practice the true HPD credible region is difficult to compute due to the high dimensional integral in equation (\ref{eq:CredibleIntegral}), motivating computationally efficient approximate techniques.
\par
Recently a conservative approximation of the HPD credible set was proposed by \citet{[10]} which exploits developments in probability concentration theory. The approximate HPD credible set $C^{\prime}_{\alpha}$ is given by:
\begin{equation}
C^{\prime}_{\alpha} := \lbrace \kappa : f(\kappa) + g(\kappa) \leq \epsilon^{\prime}_{\alpha} \rbrace,
\end{equation}
with approximate level-set threshold
\begin{equation}
\epsilon^{\prime}_{\alpha} = f(\kappa^{\text{map}}) + g(\kappa^{\text{map}}) + \tau_{\alpha} \sqrt{N} + N,
\end{equation}
where the bounding term $\tau_{\alpha} = \sqrt{16 \log(3 / \alpha)}$ which in turn is constrained to confidence $\alpha \in \big ( 4\exp(-N/3) \;, 1  \big )$. The error of this approximation is bounded above by
\begin{equation}
0 \leq \epsilon^{\prime}_{\alpha} - \epsilon_{\alpha} \leq \eta_{\alpha} \sqrt{N} + N,
\end{equation}
where $\eta_{\alpha} = \sqrt{16 \log (3/\alpha)} + \sqrt{1/\alpha}$. This upper-bound is typically conservative, meaning the isocontour is at all times larger than the true isocontour (\textit{i.e.} this estimator will never produce an underestimate). In \citet{[M2]} the bound on recovered local error bars was found to be $\pm 10$ to $15\%$ larger than the true MCMC \emdash yet could be computed $\mathcal{O}(10^6)$ times faster. A similar comparison was done by \citet{[12]} in a radio interferometric setting.
\par
The concept of approximate HPD credible regions is particularly useful as it allows us to explore high dimensional posteriors \emdash many orders of magnitude larger than \textit{state-of-the-art} MCMC techniques are currently able to accomodate \emdash in a computationally efficient manner.

\section{Bayesian peak locations} \label{sec:BayesianLocation}

Often one wishes to know the location of a feature of interest within the reconstructed convergence $\kappa^{\text{map}}$. Typically, this uncertainty is assessed \textit{via} bootstrapping of the recovered image for a large number of simulated noise fields \citep[as in \textit{e.g.}][]{[9]}.
\par
With the concept of approximate HPD credible regions in mind, we propose a novel Bayesian approach to quantifying uncertainty in the peak location which we will refer to as the \textit{`Bayesian location'}.
\par
In essence the Bayesian location is computed as follows: A feature of interest is removed from the recovered convergence map, this feature is then inserted back into the convergence map at a new position to create a surrogate convergence map, if this surrogate map is within the approximate credible set then the position at which the feature was inserted cannot be rejected, if the surrogate is not in the approximate credible set then the position can be rejected. This process is computed for a sample of the total posible insertion positions, eventually providing an isocontour of `acceptable' positions. This isocontour, at a well-defined confidence level, is the Bayesian location.

\subsection{Bayesian Location}
Suppose we recover a (MAP) convergence field $\kappa^{\text{map}}$ \textit{via} optimization of the objective function defined in equation (\ref{eq:optimization}) which contains a feature of interest (\textit{e.g.} a large peak). Let us define the sub-set of pixels which contain this feature to be $\Omega_{\mathcal{Z}} \subset \Omega$, where $\Omega$ is the entire image domain.
\par
To begin with, extract the feature $\mathcal{Z} = \kappa^{\text{map}} \mathbb{I}_{\Omega_{\mathcal{Z}}}$, \textit{i.e.} a convergence field which contains only the feature of interest. Now we adopt the process of \textit{segmentation inpainting} \citep{[11],[12],[M1]} to create a convergence field realization without the feature of interest $\mathcal{Z}$ but with background signal replaced.
\par
Mathematically segmentation inpainting is represented by the iterations
\begin{equation} \label{eq:inpainting}
\kappa^{(t+1),\text{sgt}} = \kappa^{\text{map}} \mathbb{I}_{\Omega \setminus \Omega_{\mathcal{Z}} }  + \Lambda \text{soft}_{\lambda}(\Lambda^{\dag} \kappa^{(t),\text{sgt}})\mathbb{I}_{\Omega_{\mathcal{Z}}},
\end{equation}
where $\Lambda$ is an appropriately selected dictionary \emdash for this purpose we simply use the Daubechies 8 (DB8) wavelet dictionary with 8-levels and $\lambda$ is the soft-thresholding parameter.
\par
Following the wavelet inpainting, in order to separate the true feature from the background residual convergence the signal which was inpainted into the region $\Omega_{\mathcal{Z}}$ is subtracted from the extracted feature $\mathcal{Z}$ \emdash effectively accounting for the residual background signal which would likely have been present even in the absence of the feature $\mathcal{Z}$. At this junction the surrogate convergence $\kappa^{\text{sgt}}$ is hypothesis tested for physicality \citep{[12],[M1]}.
\par
If a feature is not found to be physical, the algorithm terminates at this point as, fundamentally, it is illogical to evaluate the uncertainty in position of an object of which you cannot statistically determine the existence.
\par
Now that we have successfully isolated $\mathcal{Z}$ we can insert it back into the surrogate field $\kappa^{\text{sgt}}$ at a perturbed position. It is then sufficient to check if
\begin{equation}
f(\kappa^{\text{sgt}\prime}) + g(\kappa^{\text{sgt}\prime}) \leq \epsilon_{\alpha}^{\prime},
\end{equation}
where $\kappa^{\text{sgt}\prime}$ represents the surrogate with the feature  $\mathcal{Z}$ inserted at a perturbed location.
\par
If the inequality does hold, then the conclusion is that at $100(1-\alpha)\%$ confidence we cannot say that the feature could not be found at this location. If the equality does not hold then $\mathcal{Z}$ in its observed form could not have been found at the new location at $100(1-\alpha)\%$ confidence. The question then becomes, which perturbed positions are acceptable and which are not.
\par
With the above in mind, we propose a typical inverse nested iterative scheme to determines the pixel-space isocontour for a given feature in the reconstructed convergence field. Schematically this iterative process is outlined in Figure \ref{fig:BayesianLocationSchema}. Essentially, inverse nesting is: start in a ring 1-pixel from the MAP peak location in the first iteration, moving the ring one pixel outwards after each iteration.

\subsection{N-splitting Circular Bisection} \label{sec:N-splitting}
Inverse nested iterations are sufficiently fast for low-dimensional reconstructions $(256 \times 256)$, however as the dimensionality of the reconstructed domain grows it becomes increasingly beneficial to adopt more advanced algorithms to compute the Bayesian location in an efficient manner.
\par
We propose a novel iterative algorithm for computing the pixel-space position isocontour which we call \textit{N-splitting Circular Bisection} (N-splitting), the full details of which can be found in appendix \ref{sec:appendixa}. A brief outline of N-splitting is given below.
\par
Suppose we wish to compute positions on the Bayesian location isocontour at \textit{equiangular intervals} $\Delta \Theta$ (defined clockwise about the peak location) relative to the $y$-axis. Then we require $n = 2 \pi / \Delta \Theta$ sampling angles which are trivially given by,
\begin{equation}
\Theta_i = i \Delta \Theta,
\end{equation}
where $i$ is an iterative factor which sets the angle for a given direction $\Theta_i$.
\par
Once $\Theta_i$ is defined for a single direction, the distance $d_{\alpha}^{\prime}$ along direction $\Theta_i$ such that the objective function saturates the level-set threshold $\epsilon_{\alpha}^{\prime}$ is found by bisection. Mathematically, this is formally defined to be,
\begin{align}
d_{\alpha}^{i} &= \minT_{d} \Big \lbrace \; d \in \Gamma_i \; | \; f(\kappa_d^{\text{sgt}}) + g(\kappa_d^{\text{sgt}}) > \epsilon_{\alpha}^{\prime}   \; \Big \rbrace, \\
\Gamma_i &= \Big \lbrace q_1 \sin(\Theta_i), q_2 \cos(\Theta_i) \; | \; q_1,q_2 \in \mathbb{R}_{+} \Big \rbrace, \\
d &= \sqrt{q_1^2 + q_2^2},
\end{align}
where $\Gamma_i$ is the sub-set of the real domain which lie on the line of infinite extent along a given direction $\Theta_i$ sourced at the peak location, and $\kappa_d^{\text{sgt}}$ is the surrogate convergence map constructed by inserting the feature of interest $\mathcal{Z}$ into a perturbed location $[q_1 \sin(\Theta_i), \; q_2 \cos(\Theta_i)]$.
\par
Once a representative set of positions on the location isocontour are computed, the \textit{convex hull} is taken -- the convex hull is simply the smallest convex set which contains all samples of the location isocontour. The boundary of this closed convex set of acceptable pixels is taken as the Bayesian location.

\section{Illustrative example of the Bayesian Location} \label{sec:BayesLocationApplication}
In this section we perform sparse Bayesian reconstructions of a large cluster extracted from the Bolshoi N-body simulation \citep{[20],[6]}, upon which we construct and assess the performance of Bayesian locations for each of the four largest sub-halos. In line with previous work of \citet{[M1]} and in the related article of \citet{[6]} we refer to this extracted cluster as Bolshoi-3.
\par
We grid the Bolshoi convergence field onto a discretized complex map of dimension $(1024 \times 1024)$ so as to demonstrate that the sparse Bayesian approach can construct Bayesian estimators efficiently even when the dimension of the problem is of $\mathcal{O}(10^6)$ or larger \emdash dimensions at which MCMC techniques can become highly computationally challenging.

\begin{figure*}
	\centering
	\includegraphics[width=\textwidth, trim={0 0 0 0.75cm},clip]{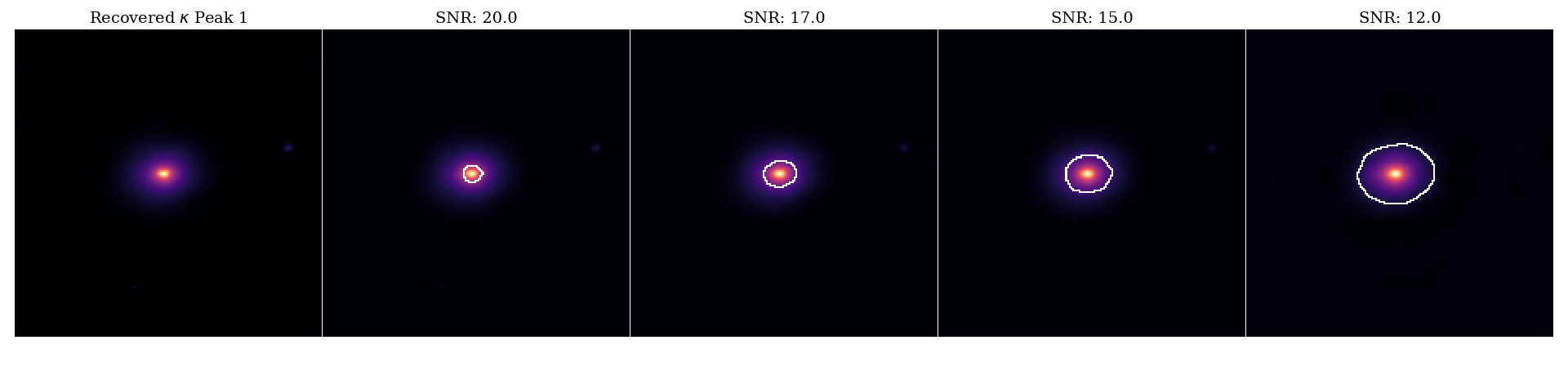}
      \put(-520,50){\large \rotatebox[origin=c]{90}{Peak 1}}
      \put(-480,115){\large True Peak}
      \put(-380,115){\large SNR: 20}
      \put(-280,115){\large SNR: 17}
      \put(-180,115){\large SNR: 15}
      \put(-80,115){\large SNR: 12} \\
    \includegraphics[width=\textwidth, trim={0 0 0 0.75cm},clip]{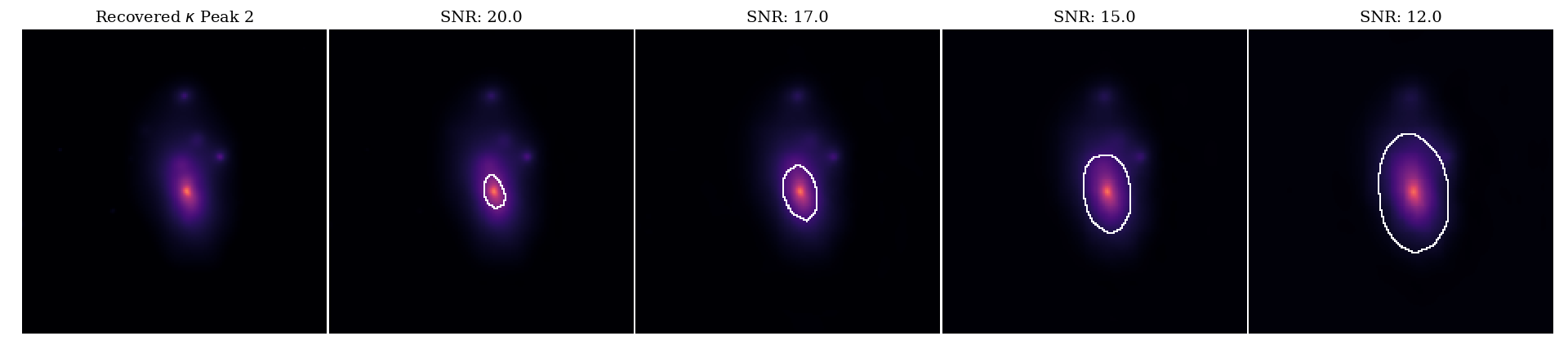}
        \put(-520,50){\large \rotatebox[origin=c]{90}{Peak 2}} \\
    \includegraphics[width=\textwidth, trim={0 0 0 0.75cm},clip]{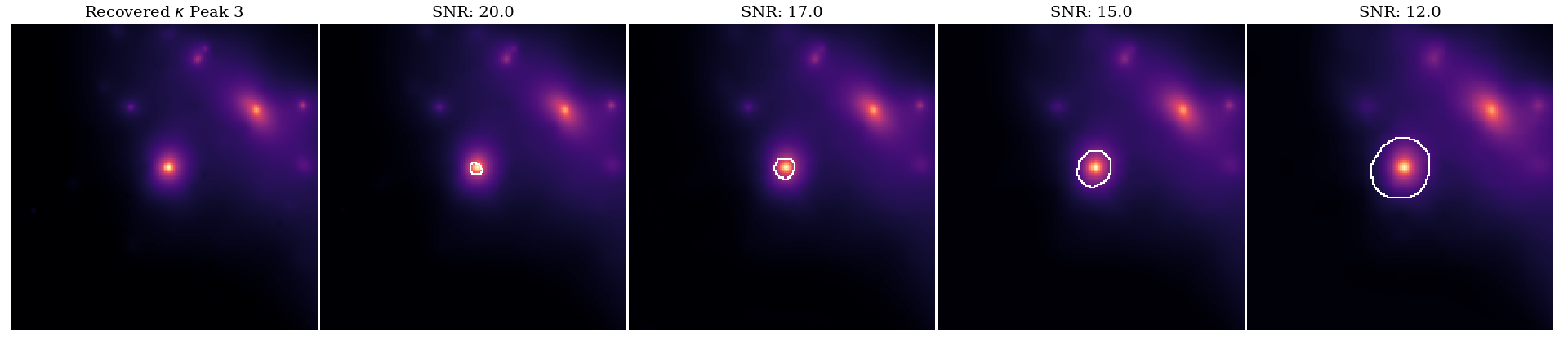}
        \put(-520,50){\large \rotatebox[origin=c]{90}{Peak 3}} \\
    \includegraphics[width=\textwidth, trim={0 0 0 0.75cm},clip]{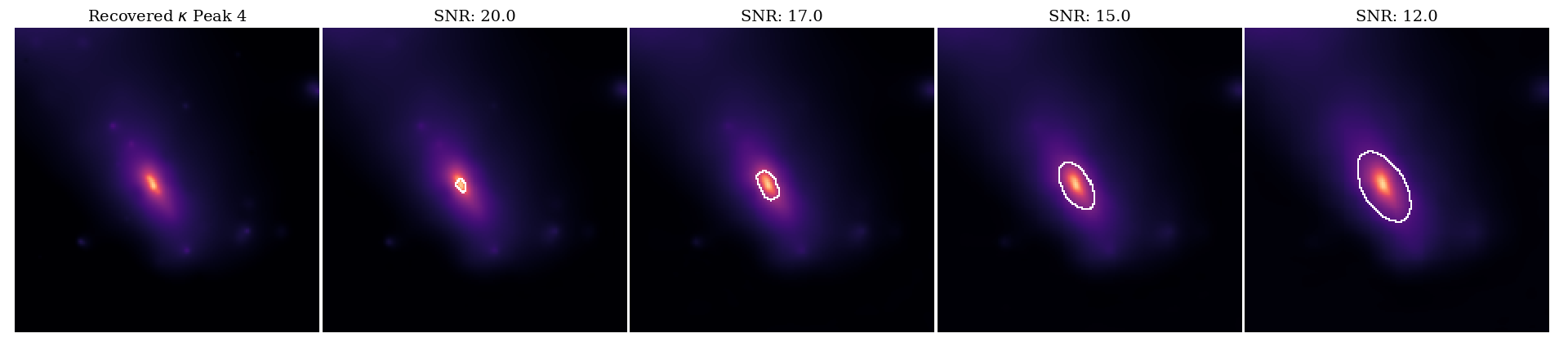}
        \put(-520,50){\large \rotatebox[origin=c]{90}{Peak 4}} \\
    \caption{\textbf{Left to right:} Sparse Bayesian reconstructions of Bolshoi-3 peaks 1 to 4  (\textit{top to bottom} respectively) followed by Bayesian locations (see section \ref{sec:BayesianLocation}) at $99\%$ confidence for input SNR of 20.0 to 12.0 dB\emdash which are overlaid on the sparse Bayesian MAP recovered convergence maps $\kappa^{\text{map}}$ at the corresponding SNR level. As the input artificial shear becomes more contaminated with noise, the relative information content decreases, and so inferred uncertainty of the reconstructed peak positions increases, as one would logically expect. Note the asymmetry in the $99\%$ isocontour, which motivates the N-splitting searching algorithm (see section \ref{sec:N-splitting} and Appendix A) rather than a naive circular inference (\textit{e.g.} finding the maximal $x$ and $y$ displacements and assuming a circular isocontour). Finally, observe that the $99\%$ isocontour for Peaks 3 and 4 are proportionally more tightly constrained than Peaks 1 and 2. This is due to the local information density typically being higher in more signal dense regions \emdash perturbations to pixels in more information dense regions are more tightly constrained and can therefore move less distance before saturating the approximate level-set threshold $\epsilon_{\alpha}^{\prime}$. This effect has been observed in the context of \textit{local credible intervals} as presented in \citet{[12]} and introduced to the weak lensing setting in \citet{[M2]}. }
    \label{fig:Bolshoi_3_peaks}
\end{figure*}
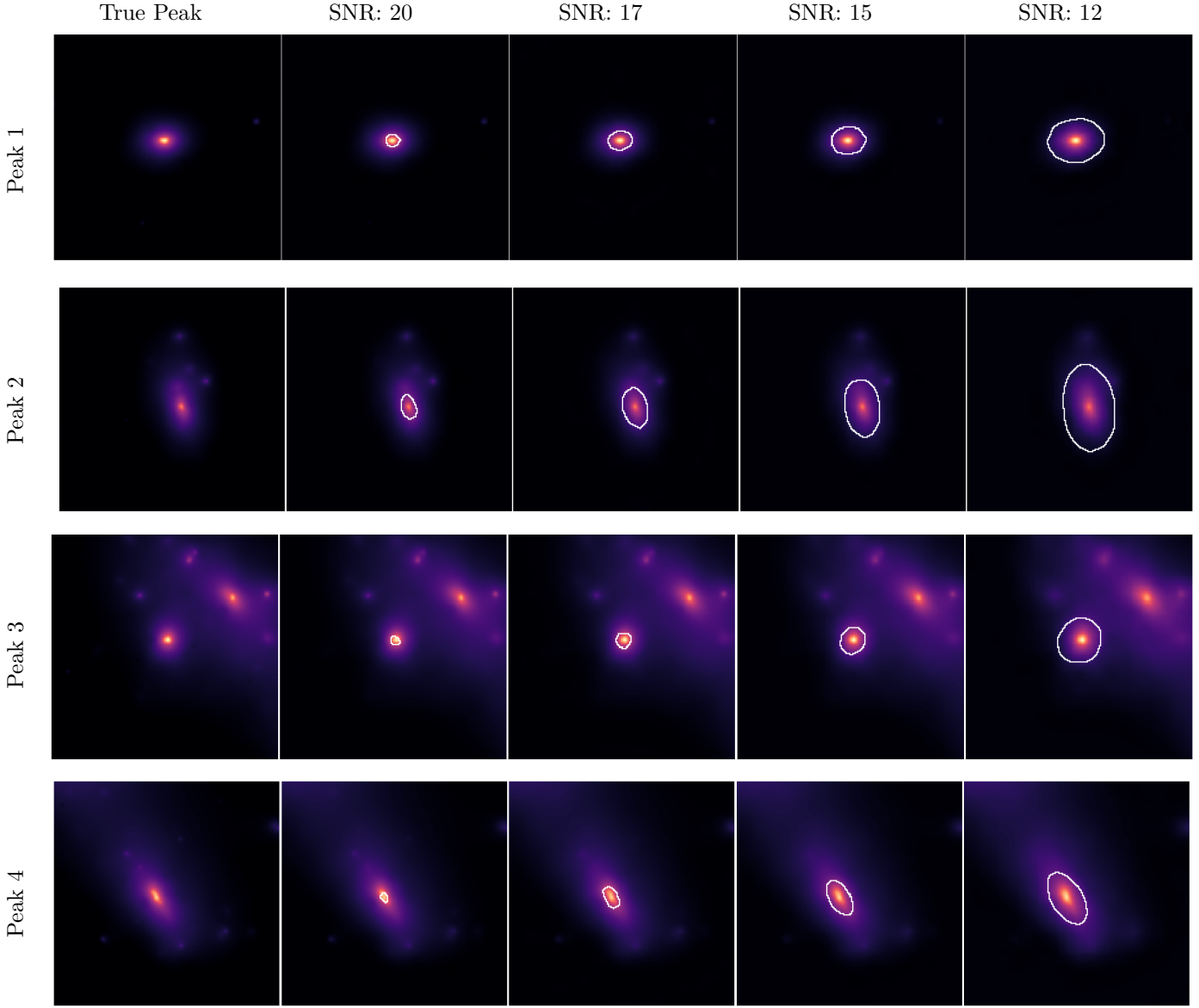

\subsection{Methodology} \label{sec:methodology}

First, we construct a complex discretized set of artificial shear measurements $\tilde{\gamma} \in \mathbb{C}^M$ by,
\begin{equation}
\tilde{\gamma} = \Phi \kappa,
\end{equation}
where $\kappa$ is the input Bolshoi-3 convergence map. We then contaminate these mock measurements with noise $n$, which for simplicity we select to be i.i.d. Gaussian noise $n \sim \mathcal{N}(0,\sigma_n^2)$ of zero mean and variance $\sigma_n^2$. The variance is selected such that the SNR of the noisy artificial shear maps can be varied, and is therefore set to be,
\begin{equation}
\sigma_n = \sqrt{\frac{\norm{\bm{\Phi} \kappa}_2^2}{N}} \times 10^{-\frac{\text{SNR}}{20}}.
\end{equation}
\par
The MAP convergence field $\kappa^{\text{map}}$ is recovered \textit{via} the sparse Bayesian mass-mapping algorithm using DB10 wavelets (10-levels), and the Bayesian location for the set of 4 peaks is constructed. For a detailed discussion of how noise levels in dB translate to physical quantities such as galaxy number density see \citet{[M1]}.
\subsection{Analysis and computational efficiency} \label{sec:analysis}
To demonstrate this uncertainty quantification technique we construct $99\%$ confidence  Bayesian locations for the 4 largest sub-halos in the Bolshoi-3 cluster, for input SNR in decibels (dB) of $\in \lbrace 12, 15, 17, 20 \rbrace$.
\par
In Figures \ref{fig:Combined_peaks} and \ref{fig:Bolshoi_3_peaks} it is apparent that, as expected, the positional uncertainty isocontour at $99\%$ confidence is smaller for less noisy data, growing in proportion to the noise. In our analysis 32 N-splitting directions (pointings) were used, though as can be seen in Figures \ref{fig:Combined_peaks} and \ref{fig:Bolshoi_3_peaks} as few as 16 directions would easily have been sufficient given the smoothness of the isocontour.
\par
Computationally, reconstruction of the MAP convergence field and computation of the Bayesian location for the complete set of peaks took $\sim 5$ minutes on a standard 2016 MacBook Air. Notably, this is a conservative \citep{[10]} and tight \citep{[M2]} approximate Bayesian inference in an over $10^6$-dimensional space on a personal laptop in minutes. Further to this, the sparse Bayesian algorithmic structure can be easily parallelizable and so this computational efficiency can be optimizerd further.

\section{Aggregate uncertainty in Peak Counts} \label{sec:peak_uncertainties}
Building on the notion of an approximate HPD credible region presented in section \ref{sec:HPD_region} we now ask the question: given a reconstructed convergence field $\kappa^{\text{map}}$, and at a given SNR threshold $K$, what is the maximum and minimum peak count at $100(1-\alpha)\%$ confidence.
\par
In this article we choose to define a \textit{peak} in $\kappa^{\text{map}}$ by a pixel $\kappa^{\text{map}}(\bm{x})$ which is larger than the 8 pixels which surround it \citep{[53]}. A point of the peak statistic is computed as follows: A threshold $K$ is taken on $\kappa^{\text{map}}$, and the \textit{peak count} (number of peaks which have intensity larger than $K$) is taken on the sub-set of pixels larger than the threshold.
\par
Formally we define the \textit{excursion set} $\Omega^{+} \subset \Omega$ as,
\begin{equation}
\Omega^{+} = \Big \lbrace \:  \bm{x} \; | \; \kappa^{\text{map}}(\bm{x}) > K \: \Big \rbrace,
\end{equation}
where $\Omega$ is the complete set of recovered pixels. Define a further sub-set $\Pi \subset \Omega^{+}$ as the set of peaks in $\Omega^{+}$:
\begin{equation} \label{eq:excursion_peak_definition}
\Pi (\kappa^{\text{map}}) = \Big \lbrace \:  \bm{x} \; | \; \kappa^{\text{map}}(\bm{x}) > \kappa^{\text{map}}(\bm{x}^{\prime}), \; \forall \; \bm{x}^{\prime} \in \mathcal{N}(\bm{x}) \: \Big \rbrace,
\end{equation}
where $\mathcal{N}(\bm{x})$ represents the set of immediately surrounding pixels.
\par
Note that this definition is not valid for pixels on the boundary of the field, and so these pixels are necessarily not considered. This not only excludes the outer edge of $\kappa^{\text{map}}$ but also any pixels surrounding masked regions (of which there are typically many).
\par
Conceptually, we would then like to know at a given threshold $K$ what is the maximum and minimum number of peaks which could exist such that the \textit{surrogate solution} $\kappa^{\text{sgt}}$ still belongs to the approximate HPD credible set $C_{\alpha}^{\prime}$.
\par
Let $\eta_{\alpha}^{\text{max}}$ be the \textit{upper bound} on the number of peaks, and $\eta_{\alpha}^{\text{min}}$ be the \textit{lower bound} on the number of peaks, for a given threshold $K$, at $100(1-\alpha)\%$ confidence. Further let $\eta$ be the number of peaks calculated from the MAP solution $\kappa^{\text{map}}$ at threshold $K$. Formally these quantities are given by,
\begin{align}
\eta \; &\equiv \; |\Pi(\kappa^{\text{map}})|, \label{eq:peak_mean} \\
\eta_{\alpha}^{\text{max}} &\equiv \maxT_{\kappa^{\text{sgt}}} \Big \lbrace \:  |\Pi(\kappa^{\text{sgt}})| \in \mathbb{R}_+ \: | \: f(\kappa^{\text{sgt}}) + g(\kappa^{\text{sgt}}) \leq \epsilon_{\alpha}^{\prime} \: \Big \rbrace, \label{eq:peak_upper} \\
\eta_{\alpha}^{\text{min}} &\equiv \minT_{\kappa^{\text{sgt}}} \Big \lbrace \:  |\Pi(\kappa^{\text{sgt}})| \in \mathbb{R}_+ \: | \: f(\kappa^{\text{sgt}}) + g(\kappa^{\text{sgt}}) \leq \epsilon_{\alpha}^{\prime} \: \Big \rbrace, \label{eq:peak_lower}
\end{align}
where $|\Pi(\kappa)|$ is the \textit{cardinality of the peak set} of a convergence field $\kappa$.
\par
It is not all obvious how to locate the extremum of optimization problems given in equations (\ref{eq:peak_upper}) and (\ref{eq:peak_lower}) as they are inherently non-linear, non-convex problems. We can, however, propose a logical iterative approach to calculate well motivated approximations to the upper and lower peak count limits $\eta_{\alpha}^{\text{max}}$ and $\eta_{\alpha}^{\text{min}}$.

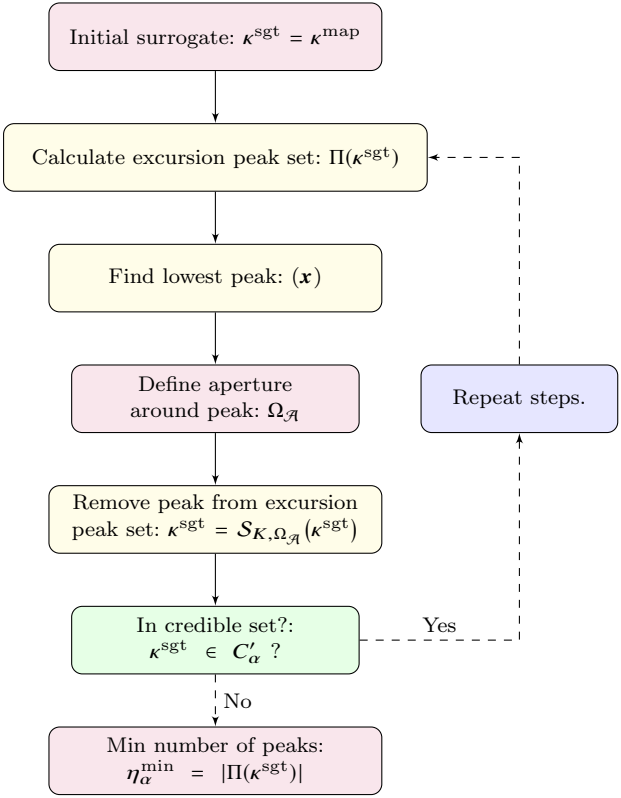
\begin{figure}
\begin{center}
\begin{tikzpicture}[node distance = 2cm, auto]
    \node [parameter_fixed, text width=14em] (0) {Initial surrogate: $\kappa^{\text{sgt}} = \kappa^{\text{map}}$};
    \node [block, below of=0, node distance=1.6cm, text width=18em] (1) {Calculate excursion peak set: $\Pi(\kappa^{\text{sgt}})$};
    \node [block, below of=1, node distance=1.6cm, text width=14em] (2) {Find lowest peak: ($\bm{x}$)};
    \node [parameter_fixed, below of=2, node distance=1.6cm] (3) {Define aperture around peak: $\Omega_{\mathcal{A}}$};
    \node [block, below of=3, node distance=1.6cm, text width=14em] (4) {Remove peak from excursion peak set: $\kappa^{\text{sgt}} = \mathcal{S}_{K, \Omega_{\mathcal{A}}} \big ( \kappa^{\text{sgt}} \big )$};
    \node [decision, below of=4, node distance=1.6cm] (5) {In credible set?: $\kappa^{\text{sgt}} \in C_{\alpha}^{\prime}$ ? };
    \node [iteration, right of=3, node distance=4.0cm, text width=8em] (6) {Repeat steps.};
    \node [parameter_fixed, below of=5, node distance=1.6cm, text width=14em] (7) {Min number of peaks: $\eta_{\alpha}^{\text{min}} = |\Pi(\kappa^{\text{sgt}})|$};

    \path [line] (0) -- (1);
    \path [line] (1) -- (2);
    \path [line] (2) -- (3);
    \path [line] (3) -- (4);
    \path [line] (4) -- (5);
    \path [line,dashed] (5) -| node[near start] {Yes}(6);
    \path [line,dashed] (6) |- (1);
    \path [line,dashed] (5) -- node{No}(7);

\end{tikzpicture}
\caption{Schematic representation of the iteration steps in finding the Bayesian lower bound $\eta_{\alpha}^{\text{min}}$ at confidence $100(1-\alpha)\%$ of the peak count $|\Pi|$ for a given MAP reconstruction $\kappa^{\text{map}}$.}
\label{fig:BayesianLowerBound}
\end{center}
\end{figure}

\subsection{Approximate Bayesian Lower Bound on Peak Counts} \label{sec:lower_peak_bound}
It is perhaps conceptually more straightforward to minimize the cardinality of the peak set and so we will first describe this process.
\par
To calculate an approximate bound on $\eta_{\alpha}^{\text{min}}$ we begin by creating the initial peak set $\Pi$ from the recovered convergence $\kappa^{\text{map}}$. The peak in $\Pi(\kappa^{\text{map}})$ with lowest magnitude is located. The shortest distance $r_{\text{min}}$ from the pixel location $\bm{x}$ to a pixel $\bm{x^{\prime}}$ such that $\kappa^{\text{map}}(\bm{x}^{\prime}) = y$ (where $y$ is some magnitude at which it is assumed the peaks influence is sufficiently small) is computed in Euclidean space as $r_{\text{min}} = | \bm{x} - \bm{x}^{\prime}|$ \emdash within this paper we simply set $y=0$.
\par
Let us define the region of interest $\Omega_{\mathcal{A}} \subset \Omega$ to be a circular aperture centered on the peak pixel location $\bm{x}$ with radius $r_{\text{min}}$. Conceptually, this acts as a proxy for the area effected by a large over-density sourced at the location of the peak.
\par
Now, define a \textit{down-scaling kernel} $\mathcal{S}_{K, \Omega_{\mathcal{A}}} \in \mathbb{C}^{N \times N}$ which has the action of scaling the magnitude of the sub-set $\kappa^{\text{map}} \mathbb{I}_{\Omega_{\mathcal{A}}}$ of pixels belonging to the region of interest $\Omega_{\mathcal{A}}$ onto $[0,K]$. Application of the down-scaling operator returns a surrogate solution $\kappa^{\text{sgt}}$. Mathematically this is,
\begin{equation} \label{eq:scaling}
\kappa^{\text{sgt}} = \mathcal{S}_{K, \Omega_{\mathcal{A}}} \big ( \kappa^{\text{map}} \big ) =  \kappa^{\text{map}} \mathbb{I}_{\Omega \setminus \Omega_{\mathcal{A}}} + \frac{K}{\max{ \big ( \kappa^{\text{map}} \mathbb{I}_{\Omega_{\mathcal{A}}} \big ) }} (\kappa^{\text{map}} \mathbb{I}_{\Omega_{\mathcal{A}}} ).
\end{equation}
\par
Application of $\mathcal{S}_{K, \Omega_{\mathcal{A}}}$ to an isolated region $\Omega_{\mathcal{A}}$ conserves the local topology of the field \emdash which is precisely what we want as it means we are making no assumptions about the halo profile around a peak. Removing a peak by application of $\mathcal{S}_{K, \Omega_{\mathcal{A}}}$ creates a surrogate solution $\kappa^{\text{sgt}}$ which is likely to minimize the increase in the objective function.
\par
As such $\mathcal{S}_{K, \Omega_{\mathcal{A}}}$ is a good strategy for excluding peaks from $\Pi(\kappa^{\text{map}})$ as it will likely maximize the number of peaks which can be removed from $\Pi(\kappa^{\text{map}})$ before the level-set threshold $\epsilon_{\alpha}^{\prime}$ is saturated. Thus, it will likely be near decision-theoretically optimal at minimizing equation (\ref{eq:peak_lower}), which is precisely what we want.
\par
A schematic of the iterative process proposed to find the Bayesian lower bound on the peak statistic can be seen in Figure \ref{fig:BayesianLowerBound}. In words, the process is as follows. Within each iteration, the lowest intensity peak within the peak set is removed forming a new surrogate convergence field $\kappa^{\text{sgt}}$, the objective function is recalculated and if the objective function is below the approximate level-set threshold $\epsilon_{\alpha}^{\prime}$ then the lowest peak within $\kappa^{\text{sgt}}$ is now removed, so on and so forth until the objective function rises above $\epsilon_{\alpha}^{\prime}$, at which the iterations are terminated and the minimum number of peaks is recovered.

\subsection{Approximate Bayesian Upper Bound on Peak Counts} \label{sec:upper_peak_bound}
Now we invert our perspective in order to approximate the maximum number of peaks which could be observed at a given threshold $K$ at $100(1-\alpha)\%$ confidence. Here we will be considering the non-linear maximization problem constructed in equation (\ref{eq:peak_upper}).
\par
First, we introduce the notion of the \textit{inclusion set} $\Omega^{-}$, conjugate to $\Omega^{+}$ such that $\Omega^{-} \cup \Omega^{+} \equiv \Omega$ and $\Omega^{-} \cap \Omega^{+} = \varnothing$,
\begin{equation}
\Omega^{-} = \Big \lbrace \:  \bm{x} \; | \; \kappa^{\text{map}}(\bm{x}) \leq K \: \Big \rbrace,
\end{equation}
With this in mind, we can now cast the maximization problem into a minimization problem analogous to that used before.
\par
We now wish to minimize the number of peaks that belong to the \textit{inclusion set} $\Omega^{-}$ which is by definition equivalent to maximizing the number of peaks which belong to the \textit{excursion set} $\Omega^{+}$ \emdash which is precisely what we want.
\par
Analogously to section \ref{sec:lower_peak_bound} to construct our approximate bound we calculate the further sub-set $\Pi^{-} \subset \Omega^{-}$ which is defined similarly to the relation in equation (\ref{eq:excursion_peak_definition}) such that,
\begin{equation} \label{eq:inclusion_peak_definition}
\Pi^{-}(\kappa^{\text{map}}) = \Big \lbrace \:  \bm{x} \; | \; \kappa^{\text{map}}(\bm{x}) > \kappa^{\text{map}}(\bm{x}^{\prime}), \; \forall \; \bm{x}^{\prime}  \in \mathcal{N}(\bm{x}) \: \Big \rbrace,
\end{equation}
\textit{i.e.} the sub-set of peaks below a threshold $K$.
\par
In contrast to section \ref{sec:lower_peak_bound} we now locate the largest peak in $\Pi^{-}$. Suppose that this peak is found at $\Pi^{-}(\bm{x})$, we now construct a circular aperture about $\bm{x}$ with radius $r_{\text{min}}$ as defined before. Let this circular aperture set of pixels be $\Omega_{\mathcal{A}} \subset \Omega$.
\par
Now we define an \textit{up-scaling kernel} $\mathcal{S}_{K, \Omega_{\mathcal{A}}}^{\dagger} \in \mathbb{C}^{N \times N}$ which has action,
\begin{equation}
\mathcal{S}_{K, \Omega_{\mathcal{A}}}^{\dagger} \big ( \kappa^{\text{map}} \big )  = \kappa^{\text{map}} \mathbb{I}_{\Omega \setminus \Omega_{\mathcal{A}}} + \frac{K + \Delta}{\max{ \big ( \kappa^{\text{map}} \mathbb{I}_{\Omega_{\mathcal{A}}} \big ) }} (\kappa^{\text{map}} \mathbb{I}_{\Omega_{\mathcal{A}}} )
\end{equation}
which is very slightly different to the down-scaling operator in the numerator of the second term. Here $\Delta$ is an infinitesimal quantity added such that the re-scaled peak within $\Omega_{\mathcal{A}}$ falls infinitesimally above the threshold $K$ and is therefore counted as a peak. In practice we set $\Delta$ to be $\sim 10^{-5}$ and find that adjusting this quantity by $\mathcal{O}(10^2)$ has negligible effect on the recovered uncertainties.
\par
\begin{figure}
\begin{center}
\begin{tikzpicture}[node distance = 2cm, auto]
    \node [parameter_fixed, text width=14em] (0) {Initial surrogate: $\kappa^{\text{sgt}} = \kappa^{\text{map}}$};
    \node [block, below of=0, node distance=1.6cm, text width=18em] (1) {Calculate inclusion peak set: $\Pi^{-}(\kappa^{\text{sgt}})$};
    \node [block, below of=1, node distance=1.6cm, text width=14em] (2) {Find highest peak: ($\bm{x}$)};
    \node [parameter_fixed, below of=2, node distance=1.6cm] (3) {Define aperture around peak: $\Omega_{\mathcal{A}}$};
    \node [block, below of=3, node distance=1.6cm, text width=14em] (4) {Add peak to excursion peak set: $\kappa^{\text{sgt}} = \mathcal{S}_{K, \Omega_{\mathcal{A}}}^{\dagger} \big ( \kappa^{\text{sgt}} \big )$};
    \node [decision, below of=4, node distance=1.6cm] (5) {In credible set?: $\kappa^{\text{sgt}} \in C_{\alpha}^{\prime}$ ? };
    \node [iteration, right of=3, node distance=4.0cm, text width=8em] (6) {Repeat steps.};
    \node [block, below of=5, node distance=1.6cm, text width=18em] (7) {Calculate excursion peak set: $\Pi(\kappa^{\text{sgt}})$};
    \node [parameter_fixed, below of=7, node distance=1.6cm, text width=14em] (8) {Max number of peaks: $\eta_{\alpha}^{\text{max}} = |\Pi(\kappa^{\text{sgt}})|$};

    \path [line] (0) -- (1);
    \path [line] (1) -- (2);
    \path [line] (2) -- (3);
    \path [line] (3) -- (4);
    \path [line] (4) -- (5);
    \path [line,dashed] (5) -| node[near start] {Yes}(6);
    \path [line,dashed] (6) |- (1);
    \path [line,dashed] (5) -- node{No}(7);
    \path [line] (7) -- (8);

\end{tikzpicture}
\caption{Schematic representation of the iteration steps in finding the Bayesian upper bound $\eta_{\alpha}^{\text{max}}$ at confidence $100(1-\alpha)\%$ of the peak count $|\Pi|$ for a given MAP reconstruction $\kappa^{\text{map}}$.}
\label{fig:BayesianUpperBound}
\end{center}
\end{figure}
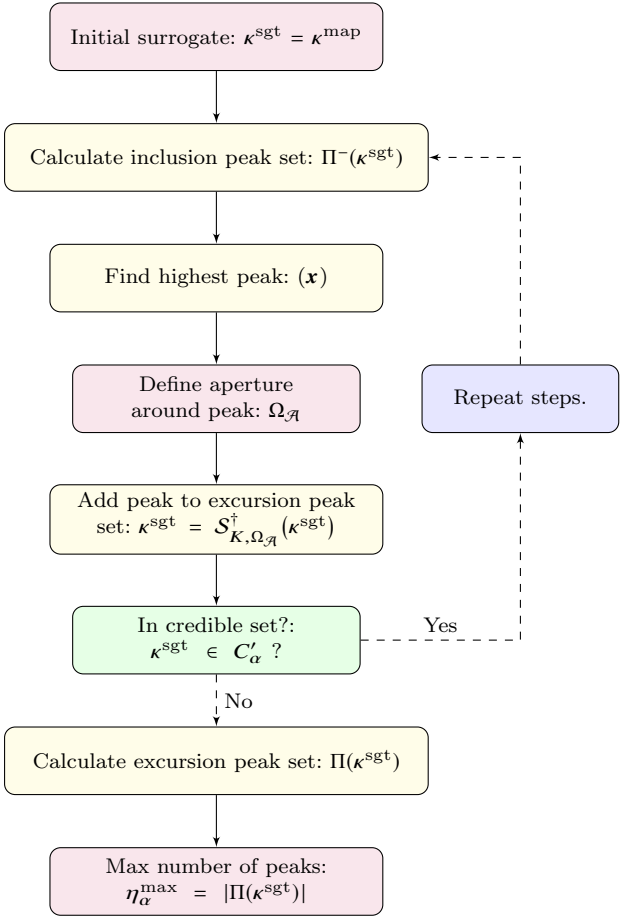

With these conceptual adjustments we then follow the same iterations discussed in section \ref{sec:lower_peak_bound} to find the approximate Bayesian upper bound on the peak count $\eta_{\alpha}^{\text{max}}$. Schematically this is given in Figure \ref{fig:BayesianUpperBound}.
\par
Finally we return the tuple $\big ( \eta_{\alpha}^{\text{min}}, \eta , \eta_{\alpha}^{\text{max}} \big )$ which is in the form $\big ($minimum, most likely, maximum$\big )$ at $100(1-\alpha)\%$ confidence.

\begin{figure}
	\centering
    \large Ground Truth Buzzard $2048 \times 2048$ Convergence $\kappa$\par
	\includegraphics[width=0.5\textwidth]{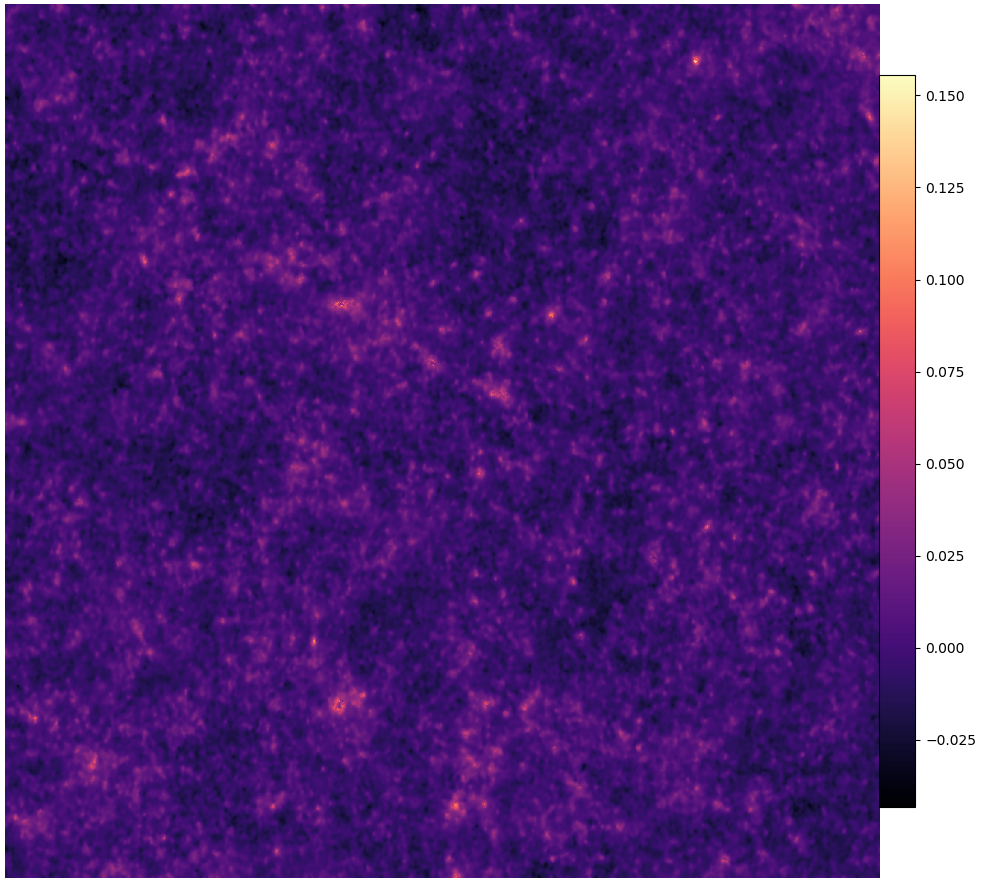}
    \caption{Input $2048 \times 2048$ convergence map extracted from the Buzzard N-body simulation.}
    \label{fig:Buzzard_input_2048}
\end{figure}

\begin{figure*}
	\centering
  \large Bayesian Uncertainty in $2048 \times 2048$ Buzzard Peak statistic: SNR = 30 dB\par
	\includegraphics[width=\textwidth, trim={0 0 0 1.2cm},clip]{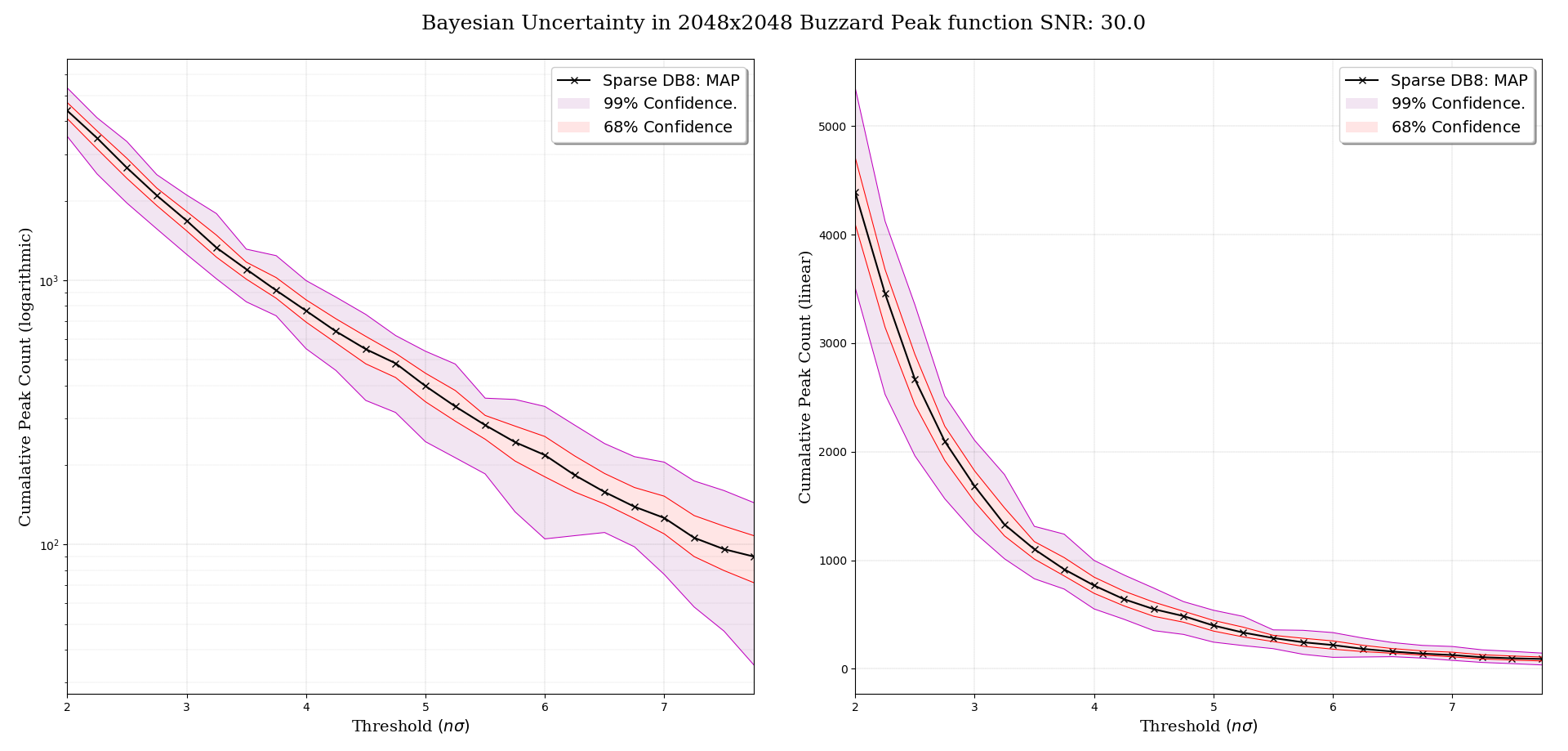}
    \caption{Cumulative peak statistic for a $2048 \times 2048$ planar convergence map extracted from the Buzzard V-1.6 simulation (see section \ref{sec:buzzard_data}) contaminated with i.i.d. Gaussian noise such that the discretized simulated shear (see section \ref{sec:methodology}) are of SNR 30 dB. The \textit{purple} outer contours are the computed upper and lower bounds at $99\%$ confidence, with the inner \textit{red} contours representing the $68\%\ (\sim 1 \sigma)$ bounds, included to aid comparison to similar literature which typically quote $1 \sigma$ errors. Note that the information content drops for higher $\sigma$ thresholds as fewer peaks are present, leading to larger relative uncertainty as fewer samples are recovered. Further note that this example is computed in a highly idealized low-noise setting.}
    \label{fig:SNR30_peak_application}
\end{figure*}

\subsection{Limitations of Re-scaling}
Suppose the SNR threshold $K$ is large enough such that during iterations in schematic of Figure \ref{fig:BayesianLowerBound} the cardinality of the excursion peak set $| \Pi(\kappa^{\text{sgt}}) |  \rightarrow 0$. In this situation even though the approximate level-set threshold $\epsilon_{\alpha}^{\prime}$ is not saturated, the algorithm is forced to stop as there are simply no more peaks to exclude (push down). At this point the strategy for removing peaks becomes locally ill-defined. Effectively this is a clipping artifact. To avoid this effect entirely, if $| \Pi(\kappa^{\text{sgt}}) | = 0$ at any point within the iterations at a given threshold, the lower bound  $\eta_{\alpha}^{\text{min}}$ at threshold $K$ is set to $0$, \textit{i.e.} we are infinitely uncertain by construction.
\par
Analogously, consider the case when $K$ is small enough that during the iterations in schematic \ref{fig:BayesianUpperBound} the cardinality of the inclusion peak set $| \Pi^{-}(\kappa^{\text{sgt}}) | \rightarrow 0$. In this situation there are simply no more peaks to include (pull up). Again we remove this clipping effect by setting $\eta_{\alpha}^{\text{max}}$ at threshold $K$ is set to $| \Pi(\kappa^{\text{sgt}}) |$.
\par
Typically these clipping effects only occur for very small $K \leq 2$ or very large $K \geq 8$ thresholds, and so a wealth of information can be extracted from the intervening scales. Low thresholds clip the upper limit $\eta_{\alpha}^{\text{max}}$ as the cardinality of the peak set drops to 0 quickly, but the objective function rises comparatively slowly, as this SNR range is statistically dominated by noise. High threshold clip the lower limit $\eta_{\alpha}^{\text{min}}$ simply due to the inherently low count of peaks at high SNR thresholds.
\par
Further to this, the decision-theory approach adopted here for locating the maximal and minimal values of the cumulative peak statistic is based on several assumption: removing lower peaks increases the objective function by less than larger peaks; the extent of a peak (dark matter over-density) is approximated by a circular aperture; and removal of a peak has little to no effect on locations in the image domain which are outside of this aperture. All three of these assumptions are very reasonable.
\par
Although further computational optimizations are not an immediate concern since our approach is already highly computationally efficient, we acknowledge that this iterative approach for removing peaks can easily be formulated as a bisection style problem which is likely to drastically reduce the computation time further \emdash particularly for low thresholds, as it mitigates the number of trivial noise peak removal recalculations which are done in the brute force approach presented above. In future, should computational efficiency become of primary interest this speed up will be considered.

\section{Illustrative example of Peak Uncertainties} \label{sec:PeakDemonstration}
In this section we apply the sparse Bayesian mass-mapping pipe-line to high resolution $(2048 \times 2048)$ convergence maps extracted from the Buzzard V-1.6 N-body\footnote{Obtained due to our affiliation with the LSST-DESC collaboration.} simulation, upon which we construct the cumulative peak statistic (number of peaks above a threshold as a function of the threshold). Additionally, we recover the $99\%$ approximate Bayesian constraints on the peak count at each threshold, from which we infer the $68\%$ constraint so as to aid the reader in comparison to typical $1 \sigma$ error-bars quoted in related literature.
\subsection{Simulated Data-sets} \label{sec:buzzard_data}
The Buzzard V-1.6 N-body simulation convergence catalog \citep{DeRose2018,wechsler2018} has a quarter sky coverage and is extracted by full ray-tracing. For wide-fields the \textit{flat sky approximation} breaks down \citep{[3]} and so this quarter sky coverage was reduced to smaller planar patches.
\par
The complete quarter sky convergence catalog was projected into a coarse HEALPix\footnote{http://healpix.sourceforge.net/documentation.php}\citep{Gorski2005} pixelisation ($N_{\text{side}} = 4$). Inside of each pixel, we further tessellated the largest square region which we then project into a $2048 \times 2048$ grid. These gridded convergence maps formed our ground truth, discretized convergence fields.
\par
As HEALPix samples in such a way as to provide equal area pixels, and the Buzzard simulation galaxy density is fairly uniform, each extracted square region contained $\sim 2 \times 10^7$ galaxies leading to $\sim 5$ galaxies per pixel.
\par
Due to a comparatively low density of samples, Poisson noise is prevalent, and thus extracted planar regions were passed through a multi-scale Poisson denoising algorithm. This consisted of a forward Anscombe transform (in order to Gaussianise the Poisson noise), several TV-norm (total-variation) denoising optimizations of differing scale, followed by an inverse Anscombe transform \citep[as in][]{[M2],[6]}. A more involved treatment could be applied, but this approach is sufficient to demonstrate our peak reconstructions.

\begin{figure*}
	\centering
  \large Bayesian Uncertainty in $2048 \times 2048$ Buzzard Peak statistic: SNR = 25 dB\par
	\includegraphics[width=\textwidth, trim={0 0 0 1.2cm},clip]{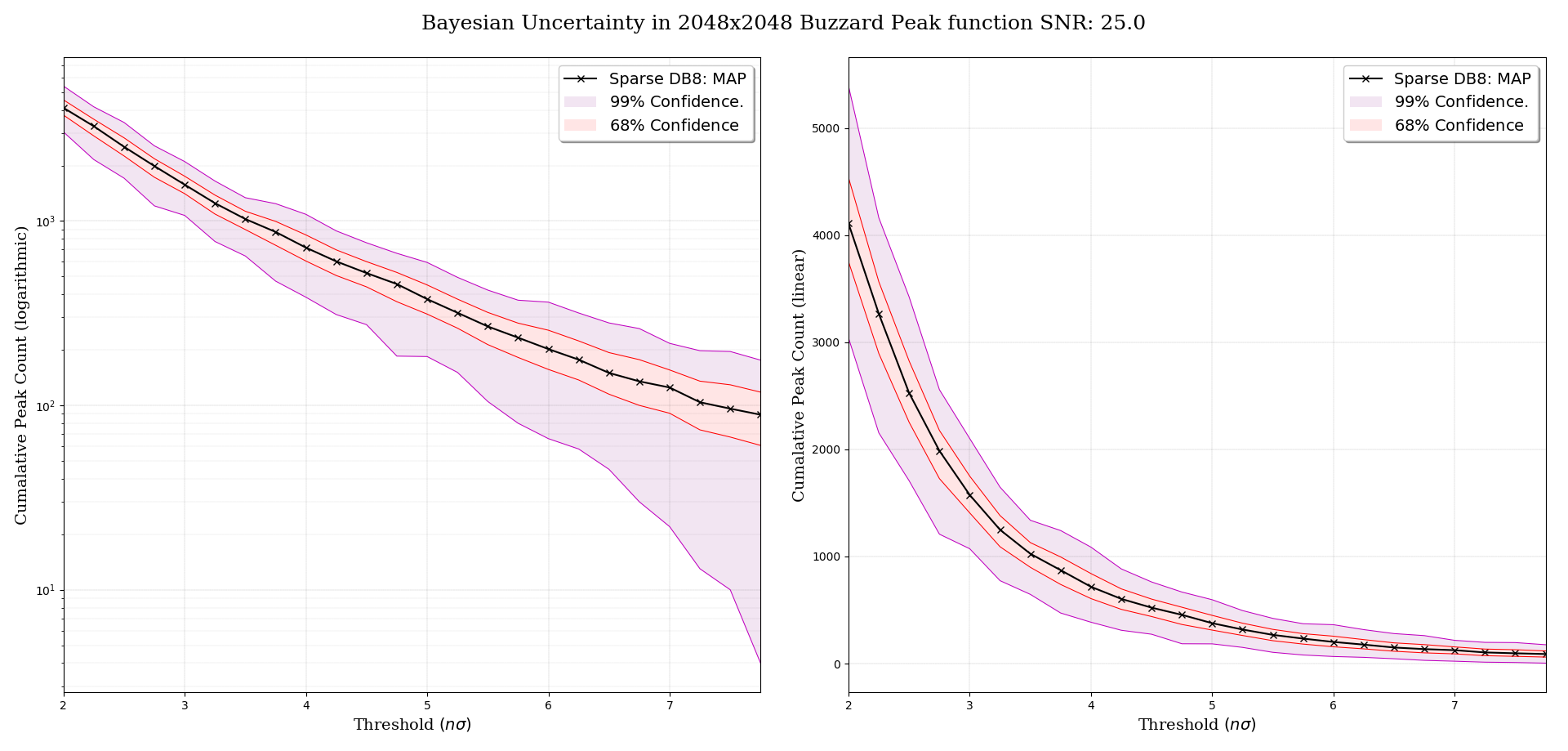}
    \caption{Cumulative peak statistic for a $2048 \times 2048$ planar convergence map extracted from the Buzzard V-1.6 simulation (see section \ref{sec:buzzard_data}) contaminated with i.i.d. Gaussian noise such that the discretized simulated shear (see section \ref{sec:methodology}) are of SNR 25 dB. The \textit{red} inner contours represent the upper and lower bounds at $68\%\ (\sim 1 \sigma)$ confidence, with the outer \textit{purple} contours representing the computed bounds at $99\%$ confidence.}
    \label{fig:SNR25_peak_application}
\end{figure*}

\subsection{Application to Buzzard V-1.6}
We select at random one of many planar patches produced for the following application. Following the methodology presented in section \ref{sec:methodology} we generate an artificial shear catalog which we then contaminate with independent and identically distributed (i.i.d.) Gaussian noise such that the SNR of mock shear measurements is 30 dB -- \textit{i.e.} an idealized noise-level simply for illustrative purposes.
\par
The MAP convergence estimator $\kappa^{\text{map}}$ is recovered from these noisy mock shear measurements \textit{via} our sparse Bayesian mass-mapping framework. From $\kappa^{\text{map}}$ we then calculate $\sigma^2 = \langle ( \kappa^{\text{map}} )^2 \rangle$ which we then use as a measure of the noise-level in the reconstructed convergence field. Implementing the uncertainty quantification technique presented in section \ref{sec:peak_uncertainties} we then construct the cumulative peak statistic for SNR thresholds $K \in [2\sigma, 8\sigma)$ at increments of $0.25 \sigma$ with upper and lower $99\%$ approximate Bayesian confidence limits.
\par
Figure \ref{fig:SNR30_peak_application} displays the recovered cumulative peak statistic in both a linear and logarithmic scale. Typically, similar figures in the literature will quote $1\sigma$ error-bars, and so for comparisons sake we convert the Bayesian $99\%$ confidence limits into the $68\%$ confidence limits which are comparable to $1\sigma$ constraints ( in Figure \ref{fig:SNR30_peak_application} we provide both confidence limits for clarity).
\par
Complete reconstruction of the peak statistics for 24 threshold bins, each with approximate Bayesian upper and lower bounds, for a $2048 \times 2048$ resolution convergence map, with DB11 (11-level) wavelets, took $\sim 2$ hours on a 2016 MacBook Air. This is a non-trivial Bayesian inference in over $4\times 10^6$ dimensions, and so $2$ hours is a very reasonable computation time \emdash though further speedups are possible, \textit{e.g.} we can trivially parallelize the calculations for each threshold leading to an increase in computational efficiency by a factor of the number of thresholds taken (in our case $24$).
\par
Additionally, the computational bottleneck is for lower thresholds as many low-intensity peaks must be removed, and thus an adaptive scheme could be implemented as discussed previously to avoid unnecessary sampling of these thresholds. With the aforementioned speed-ups, computation of the complete peak statistic is likely to take $\mathcal{O} (\text{minutes})$ on a personal laptop.
\par
\begin{figure*}
	\centering
  \large Bayesian Uncertainty in $2048 \times 2048$ Buzzard Peak statistic: SNR = 20 dB\par
	\includegraphics[width=\textwidth, trim={0 0 0 1.2cm},clip]{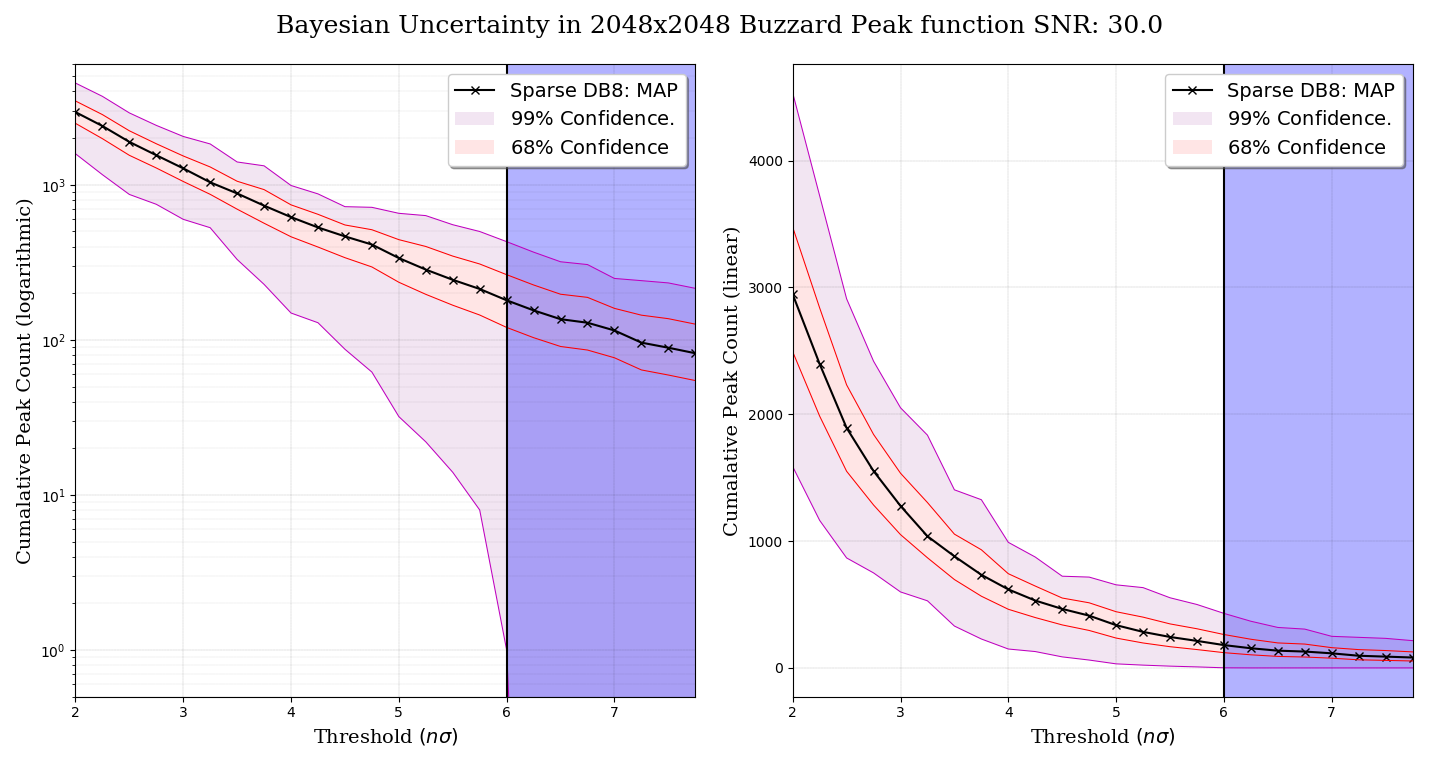}
    \caption{Cumulative peak statistic for a $2048 \times 2048$ planar convergence map extracted from the Buzzard V-1.6 simulation (see section \ref{sec:buzzard_data}) contaminated with i.i.d. Gaussian noise such that the discretized simulated shear (see section \ref{sec:methodology}) are of SNR 20 dB. The \textit{red} inner contours represent the upper and lower bounds at $68\%\ (\sim 1 \sigma)$ confidence, with the outer \textit{purple} contours representing the computed bounds at $99\%$ confidence. The shaded \textit{blue} region indicates threshold values for which at $99\%$ confidence the data cannot rule out the possibility that no peaks exist above this threshold (note that in these regions the lower bound is technically $0$ and there still exists a well defined upper bound which is given). Comparing this plot to Figure \ref{fig:SNR30_peak_application} we see that as the noise level increases the $68\%$ and $99\%$ confidence isocontours expand (as one would expect) and that in all cases the MAP peak statistics do not disagree at $99\%$ confidence.}
    \label{fig:SNR20_peak_application}
\end{figure*}

Following this initial analysis we reduce the SNR to investigate the effect of increased noise on shear measurements to the cumulative peak statistics within our Bayesian framework. We first decrease the SNR to 25 dB, seen in Figure \ref{fig:SNR25_peak_application}. Following which, we then reduce the input SNR futher to 20 dB, the corresponding results being plotting in Figure \ref{fig:SNR20_peak_application}. This higher noise level of 20 dB is still a very optimistic (somewhat unrealistic) estimate of what upcoming surveys may reach; however in this paper we are primarily focused on demonstrating the methodology and leave detailed realistic simulations and forecasting for future work. A detailed description of how these noise levels in dB translate into observation contraints (\textit{e.g.} galaxy number density \textit{e.t.c.}) can be found in \citep{[M1]}.

\subsection{Analysis of Peak statistic}
Figures \ref{fig:SNR30_peak_application}, \ref{fig:SNR25_peak_application} and \ref{fig:SNR20_peak_application} clearly show that as the noise level in the discretized complex shear field increases the isocontours of the cumulative peak statistic at $99\%$ and $68\%$ loosen noticeably. Therefore this, unsurprisingly, indicates that cleaner measurements are likely to give tighter constraints on cosmological parameters -- though it should be noted that increasing the number of data-points (\textit{i.e.} pixels) would have a similar effect to reducing the noise level per pixel.
\par
For an SNR of 20 dB (Figure \ref{fig:SNR20_peak_application}) the first feature of note is the shaded blue region which indicates that for high thresholds the lower bound on the number of peaks at $99\%$ confidence is consistent (and clipped) at 0 \emdash this is saying that at $99\%$ confidence the true number of peaks at a threshold in the blue shaded region could be 0. Note that in the blue region the Bayesian upper bound is still entirely valid, it is only the Bayesian lower bound which within our novel approach is no longer well defined.
\par
Clearly the upper and lower bounds on the peak count statistic is dependent on the threshold one is considering and the total area over which observations are made -- for wide-field surveys, more data is collected which is likely to reduce the variance of the statistic. In a general sense we summarize the mean (over all considered thresholds $K$) order of magnitude percentage spread on the peak statistic for the considered SNR thresholds below.
\par
At input SNR of 20 dB, for thresholds $\in [2 \sigma, 6 \sigma)$ on a single $2048 \times 2048$ planar patch the upper and lower bounds exist and are of $\mathcal{O}(48\%)$ at $99\%$ confidence and of $\mathcal{O}(13\%)$ at $68\%$.
\par
At input SNR of 25 dB, for thresholds $\in [2 \sigma, 8 \sigma)$ on a single $2048 \times 2048$ planar patch the upper and lower bounds exist and are of $\mathcal{O}(25\%)$ at $99\%$ confidence and of $\mathcal{O}(7\%)$ at $68\%$.
\par
At input SNR of 30 dB, for thresholds $\in [2 \sigma, 8 \sigma)$ on a single $2048 \times 2048$ planar patch the upper and lower bounds exist and are of $\mathcal{O}(15\%)$ at $99\%$ confidence and of $\mathcal{O}(3\%)$ at $68\%$.
\par
These illustrative examples imply that for the Bayesian peak statistic to tightly constrain the cumulative peak statistic comparitively larger and or cleaner data-sets may be required -- or,  of course, a more informative prior (though this must be well justified). However, to reduce the shot noise introduced \textit{via} intrinsic ellipticities more galaxies must be observed within a given pixel.
\par
One way to increase this is to simply increase the observed number density of galaxy observations, however to do so one must observe galaxies at lower magnitude (for a fixed redshift), which inherently leads to more bright distant galaxies being detected which results in galaxy blending. Hence, increasing the number density significantly above $\sim 30$ gals/arcmin$^2$ is typically quite difficult in practice.
\par
Alternatively, the pixelisation can be adjusted to ensure that the mean galaxy count per pixel is above a given threshold \emdash though for weak lensing the majority of non-Gaussian information is stored at fine-scales, which require small pixels, and so using larger pixels to reduce the noise level is sub-optimal for information extraction.
\par
Within the definition of the up and down-scaling kernels (see sections \ref{sec:lower_peak_bound} and \ref{sec:upper_peak_bound}) we define a circular aperture around a selected peak which we define to be the extent of the peak. These regions are roughly equivalent to super-pixel regions as described in \citet{[12]}. In previous work it was shown \citep{[M2]} that for local credible intervals (\textit{c.f.} pixel level error bars) the typical error in the approximate HPD credible region is of $\mathcal{O}(12.5\%)$, and is conservative \emdash note that the quoted $25\%$ mean RMSE error is split approximately equally between the upper and lower bounds, therefore this roughly corresponds to an mean error of $12.5\%$ on both. Therefore the bounds drawn on the peak static here are likely to be $\sim 12.5\%$ less tight than the true Bayesian bounds \emdash which could be formed if one were to reconstruct the $4 \times 10^6$ dimensional posterior \textit{via} MCMC.
\par
In this paper (particularly the second half) we are primarily concerned with demonstrating how one may recover principled uncertainties on aggregate statistics of the convergence map -- such as, but not limited to, the peak statistics. Hence we do not provide detailed analysis of how these Bayesian uncertainties may effect cosmological constraints derived from such statistics -- this is saved for future work. However it is worth mentioning that one could either; leverage these uncertainties to define the data covariance in a Bayesian manner (as opposed to MC which is fast but may not necessarily be fully principled, or MCMC which is $\mathcal{O}(10^6)$ times slower than our MAP approach) before then running a standard likelihood analysis ; or perform a grid search in parameter space using these uncertainties again as the data covariance. Correctly accounting the uncertainties introduced during mass-mapping has been shown to be an important consideration for the future prospects of statistics such as this \citep{Lin2018peaks}.

\section{Conclusions} \label{sec:Conclusion}
Using the sparse Bayesian mass-mapping framework previously developed \citep{[M1],[M2]} we have presented two novel Bayesian uncertainty quantification techniques which can be performed directly on weak lensing convergence maps.
\par
The first of these techniques recovers the uncertainty in the location of a feature of interest within a reconstructed convergence map \emdash \textit{e.g.} a large peak \emdash at some well defined confidence. We call this locational uncertainty the \textit{`Bayesian location'}.
\par
Additionally, for computational efficiency we develop a novel sampling scheme of the position isocontour of a given feature which we call \textit{`N-splitting circular bisection'}. We find that sampling the position isocontour in this way could be many orders of magnitude faster in high dimensions than typical inverse nesting approaches.
\par
To evaluate this technique, we perform sparse Bayesian reconstructions of $1024 \times 1024$ convergence maps extracted from Bolshoi N-body simulation datasets upon which we compute the Bayesian location of the four largest sub-halos for a range of noise-levels.
\par
The second of theses techniques quantifies the uncertainty in the cumulative peak statistic of a recovered convergence map. With this technique we can for the first time provide principled Bayesian lower and upper bounds on the number of observed peaks at a given signal to noise threshold, for a single observation, at well defined confidence.
\par
We extract $2048 \times 2048$ convergence maps from the Buzzard V-1.6 N-body simulation, upon which we calculate the cumulative peak statistic with Bayesian upper and lower bounds at $99\%$ for a range of input noise-levels.
We also provide the $68\%$ confidence bounds which we infer from the $99\%$ bounds to aid comparison to typical bootstrapping (MC) approaches.
\par
For upcoming wide-field surveys convergence reconstruction will likely be done natively on the sphere (a single collective sample) to avoid projection effects, making bootstrapping approaches difficult and at worst infeasible due to the fact that they are only asymptotically exact.
\par
Bayesian approaches require only a single set of observations to make exact inferences, and so extend trivially to the more complex spherical setting. Moreover the novel uncertainty quantification techniques presented in this paper and those presented previously in \citet{[M1],[M2],[12]} can be rapidly computed and support algorithmic structure which can be highly parallelized, making them the ideal tools for principled analysis of convergence maps.

\section*{Acknowledgements} \label{sec:Acknowledgements}
Author contributions are summarised as follows.
MAP: conceptualisation, methodology, data curation, investigation, software, visualisation, writing - original draft;
JDM: conceptualisation, methodology, project administration, supervision, writing - review \& editing;
XC: methodology, investigation, writing - review \& editing;
TDK: methodology, supervision, writing - review \& editing.

This paper has undergone internal review in the LSST Dark Energy Science Collaboration. The internal reviewers were Chihway  Chang, Tim Eifler, and François Lanusse.
MAP is supported by the Science and Technology Facilities Council (STFC).  TDK is supported by a Royal Society University Research Fellowship (URF).  This work was also supported by the Engineering and Physical Sciences Research Council (EPSRC) through grant EP/M0110891 and by the Leverhulme Trust.
The DESC acknowledges ongoing support from the Institut National de Physique Nucl\'eaire et de Physique des Particules in France; the Science \& Technology Facilities Council in the United Kingdom; and the Department of Energy, the National Science Foundation, and the LSST Corporation in the United States.  DESC uses resources of the IN2P3 Computing Center (CC-IN2P3--Lyon/Villeurbanne - France) funded by the Centre National de la Recherche Scientifique; the National Energy Research Scientific Computing Center, a DOE Office of Science User Facility supported by the Office of Science of the U.S.\ Department of Energy under Contract No.\ DE-AC02-05CH11231; STFC DiRAC HPC Facilities, funded by UK BIS National E-infrastructure capital grants; and the UK particle physics grid, supported by the GridPP Collaboration.  This work was performed in part under DOE Contract DE-AC02-76SF00515.

\begin{figure*}
	\centering
	\includegraphics[width=0.75\textwidth]{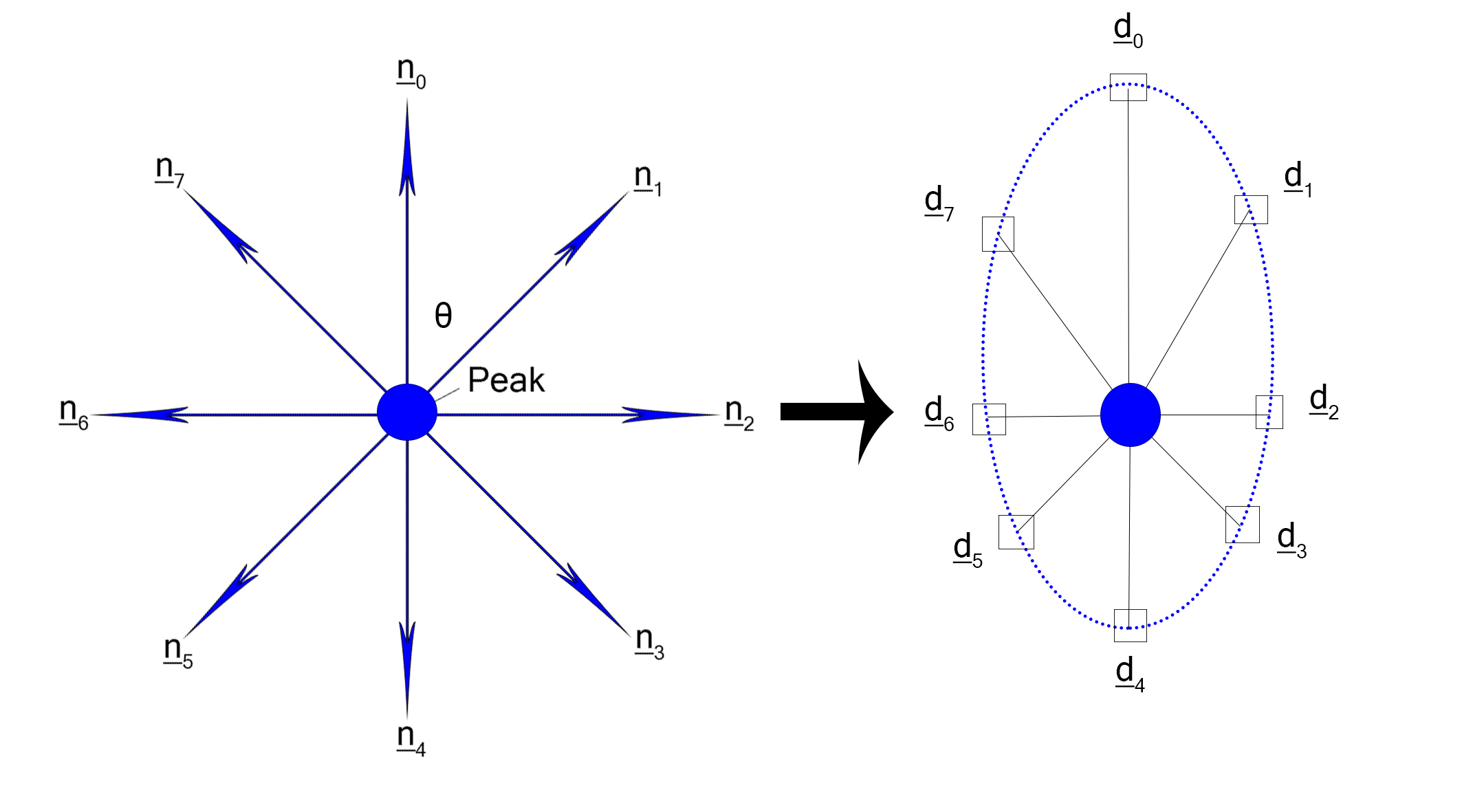}
    \caption{Representation of how the problem is broken up in N-splitting circular bisection. First the $\mathit{\textsf{n}}_i$ directions are specified (\textit{left}) at equiangular separations $\theta$ about the peak location (\textit{blue ball}). Bisection iterations are conducted as in equation (\ref{eq:circular_bisection}) along each of the directions, recovering a set of samples $\mathit{\textsf{d}}_i$ of the Bayesian location isocontour at $100(1-\alpha)\%$ confidence (\textit{right}). Provided a sufficient number of samples are taken, this boundary will fully represent the isocontour. We find typically $\approx 16$ samples are needed for $512 \times 512$ convergence reconstructions though more or less may be needed depending on the resolution and application.}
    \label{fig:n_splitting_schematic}
\end{figure*}



\bibliographystyle{mnras}
\bibliography{Refs/references.bib}

\appendix

\section{N-splitting Circular Bisection Details} \label{sec:appendixa}
In this appendix we consider the N-splitting Circular Bisection (N-splitting) algorithm for iteratively sampling the Bayesian $100(1-\alpha)\%$ confidence isocontour of the position of a feature in a reconstructed convergence map \emdash or the \textit{Bayesian Location} at $100(1-\alpha)\%$ confidence.
\par
As in the text, we begin by defining the number of directions to sample $n_T$ from which we then form the angular increment $\Delta \Theta = 2\pi / n_T$. Starting from a vector $\mathit{\bm{\mathsf{n}}}_0$ oriented along the positive $y$-axis define the $(i+1)^{\text{th}}$ pointing to be the vector,
\begin{equation}
\mathit{\bm{\mathsf{n}}}_{i+1} = \mathcal{R}_{\Delta \Theta} \mathit{\bm{\mathsf{n}}}_{i}, \quad \text{where} \quad i \in (1,n_T),
\end{equation}
and where $\mathcal{R}_{\Delta \Theta}$ is rotation by angle $\Delta \Theta$ clockwise on 2D Euclidean space \emdash a irreducible representation of which is the standard clockwise rotation matrix,
\begin{equation}
\mathcal{R}_{\theta}=
  \begin{bmatrix}
    \cos(\Delta \Theta) & \sin(\Delta \Theta) \\
    -\sin(\Delta \Theta) & \cos(\Delta \Theta)
  \end{bmatrix}.
\end{equation}
\par
Now we know the direction along which we wish to sample we construct the $(i+1)^{\text{th}}$ bisection problem which is
\begin{equation} \label{eq:circular_bisection}
d_{\alpha}^{i+1} = \minT_{d} \Big \lbrace \; d \in \Gamma_{i+1} \; | \; f(\kappa_d^{\text{sgt}}) + g(\kappa_d^{\text{sgt}}) > \epsilon_{\alpha}^{\prime}   \; \Big \rbrace,
\end{equation}
where $\kappa_d^{\text{sgt}}$ is a surrogate convergence map with the feature of interest inserted into perturbed location $d\mathit{\bm{\mathsf{n}}}_{i+1}$ and $\Gamma_{i+1}$ is sub-set of the real domain which lie on the directional line centered at the original peak location with unit vector $\mathit{\bm{\mathsf{n}}}_{i+1}$ \textit{i.e.}
\begin{equation}
\Gamma_{i+1} = \Big \lbrace a \mathit{\bm{\mathsf{n}}}_{i+1} \; | \; a  \in \mathbb{R}_{+} \Big \rbrace.
\end{equation}
A pictoral representation of how the problem is set up is provided in Figure \ref{fig:n_splitting_schematic}.
\par
For bisection we must first make an initial guess $d_0$ which we define to be square root of the number of pixels contained within the mask, as this is a typical measure of the length of a masked region. This choice is particularly logical as, if a feature of interest can be removed entirely from its masked location without saturating the level-set threshold $\epsilon_{\alpha}^{\prime}$ then it by definition must be inconclusive, \textit{i.e.} the data is insufficient evidence to say that the peak is physical.
\par
To optimize the convergence of this algorithm further (for high sampling rates, low angular increments $\Delta \Theta \leq \pi/4$) we also propagate information between pointing's. For bisection problems associated with pointing $i > 1$ the initial guess is now set to be twice the previous optimal value $d_{\alpha}^{\prime,i}$. This increases the computational efficiency by $\sim 20 \%$, in most cases.
\par
Propagating information in this way relies on the assumption that the isocontour we are searching for is somewhat smooth and continuous, which is the case for most convergence reconstructions. If there is uncertainty as to the smoothness of the isocontour it is recommended that information is not propagated and the number of pointings is increased to correctly map the isocontour structure.

\subsection{Convergence Properties}
Standard inverse nesting algorithms iteratively sample the entire sub-space of the reconstructed domain bounded by the isocontour at $100(1-\alpha)\%$ confidence, making them inefficient when one is only interested in the boundary.
\par
Consider the case where the isocontour of a reconstructed $512 \times 512$ convergence map is a circular region of radius $R$. Here inverse nesting will have to sample a square region out to $R$, which is to say the total number of samples $T_{\text{nest}}$ will at least be $R^2-1$, where 1 is removed for the central location.
\par
For our N-splitting algorithm we define $n_T$ pointings, and assume that the isocontour is relatively smooth. As the first bisection problem $\mathit{\bm{\mathsf{n}}}_0$ makes a large first guess it typically takes $4-5$ iterations to converge with a single pixel accuracy. The subsequent $n_T-1$ bisection problems converge within $3-4$ iterations. Therefore the total number of calculations $T_{\text{N-split}}$ is conservatively,
\begin{equation}
T_{\text{N-split}} = 5 + 4 (n_T-1),
\end{equation}
which is essentially independent from $R$. There is in fact a small inverse dependence which is incorporated in the number of iterations needed for convergence, though this dependence is found to be small.
\par
Comparing the computational efficiency of the two algorithms $\mathbb{E}_{512}$ where,
\begin{equation}
\mathbb{E}_{512} \equiv \frac{T_{\text{N-split}}}{T_{\text{nest}}} = \frac{5 + 4 (n_T-1)}{R^2}.
\end{equation}
Typically, we find an angular separation between pointings of $\pi/4$ (\textit{i.e.} 16 pointings) is sufficient to accurately recover the isocontour. Additionally, the circular radius is typically $15-30$ pixels which indicates that,
\begin{equation}
\frac{5 + 4 \times 15}{30^2} = 0.072 \leq \mathbb{E}_{512} \leq \frac{5 + 4 \times 15}{15^2} = 0.289,
\end{equation}
\textit{i.e.} N-splitting circular bisection on $512 \times 512$ dimensional reconstructions is $\sim 4 - 14$ times faster than inverse nesting.
\par
However, in the future we will be interested in recovering high dimensional $2048 \times 2048$ convergence maps. In this setting the number of iterations for N-splitting to converge is assumed to change by 1-2, and the number of pointings to faithfully recover the isocontour will be increase by a factor of $\sim 2$. Additionally, the radius of the circle $R$ increases by a factor of 4. Thus,
\begin{equation}
\frac{5 + 4 \times 31}{120^2} = 0.009 \leq \mathbb{E}_{2048} \leq \frac{5 + 4 \times 31}{60^2} = 0.0360,
\end{equation}
\textit{i.e.} the conservative increase in computational efficiency of N-splitting over inverse nesting for $2048 \times 2048$ becomes a factor of $\approx 30 - 112$.
\par
Further optimizations are possible, such as trivially parallelizing the bisection problems of each pointing. Doing so removes the scaling with the number of pointings, but now information about starting positions cannot be propagated.

\bsp	
\label{lastpage}
\end{document}